\documentclass[aps,prx,twocolumn,superscriptaddress,longbibliography]{revtex4-1}

\usepackage[margin=1in]{geometry}

\usepackage{amsmath}
\usepackage{amsthm,amssymb,amsfonts}
\usepackage{mathrsfs}  
\usepackage{stmaryrd}
\usepackage{accents}
\usepackage{bm}
\usepackage[mathscr]{euscript}
\usepackage[linkcolor=blue,colorlinks=true]{hyperref}
\usepackage{graphicx}
\usepackage[caption=false]{subfig}
\graphicspath{
    {figures/}
}

\newcommand{\bC}{\mathbb{C}}

\newcommand{\bP}{\mathbb{P}}
\newcommand{\bR}{\mathbb{R}}
\newcommand{\bS}{\mathbb{S}}

\newcommand{\cT}{\mathcal{T}}

\newcommand{\bra}{\langle}
\newcommand{\ket}{\rangle}
\newcommand{\llb}{\llbracket}
\newcommand{\rrb}{\rrbracket}

\newcommand{\vpi}{\varpi}
\newcommand{\vphi}{\varphi}

\newcommand{\hatb}[1]{\hat{\bm{#1}}}

\newcommand{\rvec}{\vec{r}\mkern2mu\vphantom{r}}
\newcommand{\ts}{\textsuperscript}
\DeclareRobustCommand{\ubar}[1]{\underaccent{\bar}{#1}}
\DeclareRobustCommand{\utilde}[1]{\underaccent{\tilde}{#1}}

\DeclareMathOperator{\sign}{sign}
\DeclareMathOperator{\tr}{tr}
\begin{document}


\title{{Strain-induced time reversal breaking and half quantum vortices near a putative superconducting  tetra-critical point in Sr$_2$RuO$_4$}}
\author{Andrew C. Yuan}
\affiliation{Department of Physics, Stanford University, Stanford, CA 93405, USA}
\author{Erez Berg}
\affiliation{Department of Condensed Matter Physics, Weizmann Institute of Science, Rehovot 7610001, Israel}
\author{Steven A. Kivelson}
\affiliation{Department of Physics, Stanford University, Stanford, CA 93405, USA}

\begin{abstract}
  It has been shown \cite{kivelson2020proposal} that many seemingly contradictory experimental findings concerning the superconducting state in Sr$_2$RuO$_4$ can be accounted for as resulting from the existence of an assumed tetra-critical point at near ambient pressure at which $d_{x^2-y^2}$ and $g_{xy(x^2-y^2)}$ superconducting states are degenerate.
  We perform both a Landau-Ginzburg and a microscopic mean-field analysis of the effect of spatially varying strain on such a state.
  In the presence of finite $xy$ shear strain, the superconducting state consists of two possible symmetry-related time-reversal symmetry (TRS) preserving states: $d \pm g$.
  However, at  domain walls between two such regions, TRS can be broken, resulting in a $d+ig$ state. More generally, we find that various natural patterns of spatially varying strain  induce a rich variety of superconducting textures, including half-quantum fluxoids.
  These results may resolve some of the apparent inconsistencies between the theoretical proposal and various  experimental observations,  including the suggestive evidence of half-quantum vortices \cite{jang2011observation}.
\end{abstract}
\maketitle

\section{Introduction}

\begin{figure}
\centering
\subfloat[\label{fig:isoT}]{%
  \centering
  \includegraphics[width=.5\columnwidth]{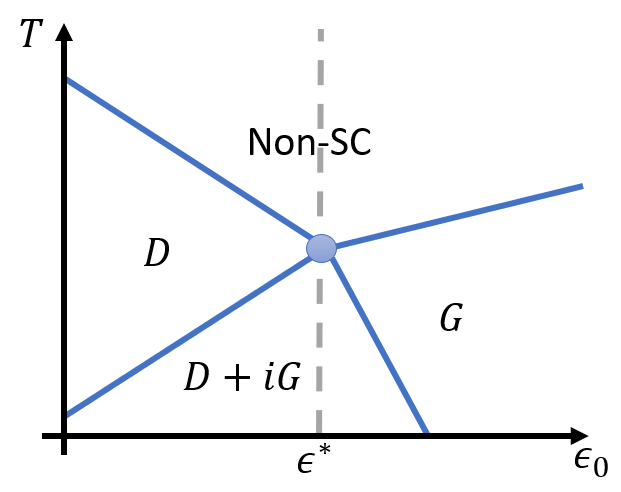}
}
\subfloat[\label{fig:shearT}]{%
  \centering
  \includegraphics[width=.5\columnwidth]{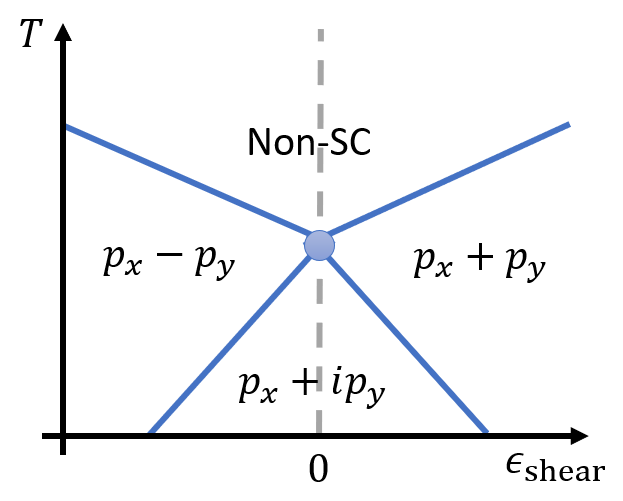}
}\hfill
\caption{
(a). Schematic phase diagram as a function of two parameters - taken here to be isotropic strain ($\epsilon_0$) and $T$, in the neighborhood of a tetra-critical point at which the transition temperatures to SC states with $d$-wave (i.e. B$_{1g}$) and  $g$-wave (i.e. A$_{2g}$) coincide.
(b). A similar schematic phase diagram -- now where the $x$-axis signifies symmetry breaking shear strain, $\epsilon_\text{shear}$
-- for a system which at zero strain favors a  SC order parameter the transforms according to a two dimensional irrep., $p_x$ and $p_y$ (i.e. $E_u$ symmetry).}

\end{figure}

If it happens that superconducting (SC) orders with two distinct symmetries are comparably favorable for some microscopic reason, it is possible to have a two-parameter phase diagram (e.g. $T$ and isotropic strain, $\epsilon_0$) that exhibits a multicritical point (e.g. at $\epsilon=\epsilon^\star$ and $T=T^\star$) at which the transition temperatures, $T_c$, of the two different orders coincide, as shown in Fig. \eqref{fig:isoT}.
For instance, a change from $s_\pm$ to $d$-wave pairing is thought to occur as a function of  doping in certain Fe-based superconductors \cite{ronny}, and it was recently conjectured  that the layered perovskite Sr$_2$RuO$_4$ (SRO) \cite{mackenzie2017even} under ambient conditions is ``accidentally'' close to such a multi-critical point 
involving either $d_{x^2-y^2}$ and $g$-wave \cite{kivelson2020proposal} or $d_{xy}$ and $s$-wave \cite{romer2020theory} pairing.
Even though both orders by themselves transform as one dimensional irreducible representations (irreps) of the point-group symmetries, near  such a multi-critical point the system can exhibit a variety of features usually associated with multi-component SC order  that transform according to a higher dimensional irrep (e.g. a $p$-wave).
Conversely, for a SC order parameter that transforms according to a 2d irrep, the point of zero shear-strain, $\epsilon_\text{shear}=0$,  can be viewed as a special case of such a multi-critical point in the $T$-$\epsilon_\text{shear}$ plane, as shown in Fig. \eqref{fig:shearT}.   
In both cases,  the response of the different components of the SC order parameter to specific components of the strain tensor can produce a variety of novel effects.

Specifically, the proximity of a multicritical point implies that even small amplitude spatial variations of the strain field can locally stabilize different distinct forms of superconducting order in different domains.
In this paper, we treat 
the case of a tetracritical point involving $d_{x^2-y^2}$ (B$_{1g}$) and $g_{xy(x^2-y^2)}$-wave (A$_{2g}$) pairing channels, where, as in Fig. \eqref{fig:isoT}, for the uniform case the coexistence regime is a $d+ig$ SC with spontaneously broken time-reversal-symmetry (TRS).
We study this problem in the mean-field approximation, both from an effective field theory (Landau-Ginzburg/non-linear sigma model) and a microscopic perspective.
Following a similar line of reasoning as in Ref. \cite{schmalian} (where TRS breaking near dislocations was investigated),
we show that  inhomogeneous strain can lead 
to a highly inhomogeneous SC state in which TRS breaking is strongly manifest only along a network of domain-walls separating regions in which the local strain favors one or another TRS preserving combination of the $d$ and $g$ order parameters.
However, this only occurs when the unstrained system is close enough to the tetra-critical point - in a sense that we make precise.  We also show that  appropriate strain patterns can induce
an order parameter texture with a spontaneous fractional magnetic flux 
that equals a half superconducting flux quantum in under a range of circumstances.

The results we obtain are quite general as they follow largely from 
symmetry considerations. 
As an application of these ideas, we explore their implications for the still vexed problem of settling the symmetry of the superconducting state in 
SRO \cite{mackenzie2017even}.
Here, there are a number of  proposals, each of which can plausibly account for a subset of the experimental observations.
Triplet pairing of any sort has seemingly been ruled out by recent NMR experiments \cite{pustogow2019constraints, ishida2020reduction, chronister2020evidence}.
Moreover, recent ultra-sound measurements \cite{ghosh2021thermodynamic, benhabib2021ultrasound},  taken at face value,  require a two-component order parameter arising from the system in the absence of strain being accidentally tuned to  a multi-critical point involving  $d_{xy}$ \& $s$ or $d_{x^2-y^2}$ \& $g$ wave pairing. However, there are at least two sets of phenomena that seemingly   challenge these  theoretical proposals:

\begin{itemize}
\item
Below a critical temperature, $T_\text{trsb}$, the SC phase appears to break time-reversal symmetry \cite{luke1998time,xia2006high,kidwingira2006dynamical} where  at zero shear strain $T_\text{trsb} \approx T_c$, while in the presence of substantial B$_{1g}$ shear strain these two transitions are split, such that $T_\text{trsb} < T_c$.
However, specific heat measurements under the same circumstances show no signature of the 
TRS breaking transition \cite{grinenko2021split, li2020heat}.
An additional constraint on theory is the recent observation \cite{newhicks} that $T_c$ can be depressed with the application of hydrostatic pressure (i.e. compressive uniform strain $\epsilon_0<0$) without producing a detectable splitting between $T_c$ and $T_\text{trsb}$ - an observation that was declared to rule out any theory based on an accidental degeneracy between two symmetry-distinct superconducting orders.

\item
A somewhat complicated experiment on mesoscale crystals adduced evidence of the existence of a topological excitation capable of admitting a half-quantum of magnetic flux \cite{jang2011observation}. 
This has been argued \cite{leggett2020symmetry} to constitute direct evidence that the SC state is a chiral $p+ip$ state, despite the contrary evidence from the NMR studies \cite{pustogow2019constraints,ishida2020reduction, chronister2020evidence}.
Although  additional evidence of half quantum vortices has been recently reported \cite{liuhalf}, it is still possible that there is an alternative explanation for the observation that does not involve fractional vortices.
However, taking the result at face value presents us with the need to identify a route to fractional vortices in a singlet SC.
\end{itemize}

In this context, our present results provide, as a matter of principle, possible routes to reconcile both these observations with the conjectured theoretical scenario.
\begin{itemize}
    \item  Despite the fact that the crystals involved in these experiments are paragons of crystalline perfection, the only plausible interpretation of the specific heat data is that the TRS breaking involves a small fraction of the electronic degrees of freedom.  This can be naturally accounted for by the theoretically expected extreme sensitivity of the  SC order to local strain, and the fact that the TRS breaking $d+ig$ order arises only in a network of domain-wall-like regions at which $|\epsilon_0-\epsilon^\star|$ and $|\epsilon_\text{shear}|$ are vanishingly small.  Moreover, so long as the typical magnitude of the inhomogeneous strain is larger than an applied strain, no significant splitting between $T_c$ and $T_\text{trsb}$ is expected.
    \item
    The fact that a half-quantum vortex can be the ground-state in the presence of a suitable strain texture similarly opens the possibility that the experimental evidence of fractional vortices in SRO is likewise consistent with the propsed scenario.
\end{itemize}

\section{Ginzburg-Landau Theory}
\subsection{Setup}
We consider a Ginzburg-Landau free energy density \cite{kivelson2020proposal} with order parameter (OP) fields $\Delta^T =(D,G)$ in the presence of an external magnetic field $\vec{B} =\vec{\nabla}\times \vec{A}$, given by
\begin{align}
  \label{eq:GL-vectorized}
  F &= V_2 +V_4 +K+ \frac{B^2}{2}\\
  V_2 &= \alpha_0 \Delta^\dagger\Delta +\Delta^\dagger \left(\bm{\alpha}\cdot \bm{\tau}\right) \Delta \nonumber\\
  V_4 &= \frac{1}{2} [\Delta^\dagger\Delta]^2 +\frac{\beta_1}{2} [\Delta^\dagger \tau_1\Delta]^2+\frac{\beta_3}{2} [\Delta^\dagger \tau_3\Delta]^2  \nonumber \\
  &+\frac{\beta_3'}{2} [\Delta^\dagger \tau_3\Delta][\Delta^\dagger\Delta] \nonumber \\
  K &=\frac{\kappa}{2} |(-i\vec\nabla -\vec A)\Delta|^2 \nonumber \\
  &+ \frac{ \kappa^\prime}{2}\left\{\left[ (i\partial_x - A_x)\Delta^\dagger\right]\tau_1\left[ (-i\partial_y - A_y)\Delta\right] + \text{c.c.}\right\} \nonumber
\end{align}
Where $\bm{\tau}$ is the vector of Pauli matrices.

The complex scalar fields $D,G$ are normalized so that the stiffness constants are equal, i.e., $\kappa_d =\kappa_g\equiv \kappa$, and that quartic isotropic coupling constant $\beta_0=1$.
We have used units such that the Cooper pair charge $2e=1$.
The quantity $\bm{\alpha}=(\alpha_1,0,\alpha_3)$ 
represents the effect of local strain as a two-component vector, where $\alpha_3 =\alpha_3(x,y)$ is proportional to the deviation of the isotropic (A$_{1g}$) strain from its critical value $\epsilon_0^\star$ and $\alpha_1 =\alpha_1(x,y)$ is proportional to the shear (B$_{2g}$, i.e., $xy$) strain.
The free energy at $\vec A=\vec 0$ respects TRS, has a $U(1)$ symmetry associated with the overall superconducting phase and, for $\alpha_1=0$,  a $D_{4h}$ point group symmetry that involves both space and the order parameters.  For example, under a rotation by $\pi/2$, $x \to y$ and $y \to - x$, $D \to -D$ and $G \to G$.
The quartic terms determine the favored form of the ordered state:
$\beta_3>0$ favors coexistence of non-zero $d$ and $g$ pairing and  $\beta_1>0$ favors the TRS breaking combinations $d\pm ig$.
The microscopic BdG calculations reported  in Sec. \eqref{sec:uniform-states} yield $\beta_1$ \& $\beta_3 >0$.

\subsection{Non-linear sigma model}

In the special case, $\alpha_j=\beta_j=0$ for all $j>0$ and $\beta'_3 =0,\kappa^\prime=0$, the free energy has 
a global $SU(2)$ symmetry that relates the two components of the order parameter.
Not too close to $T_c$, and to the extent that this symmetry is not too strongly broken, the variations of the magnitude of the order parameter are  unimportant.  This allows us to associate the relevant order parameter values with  a point on a Bloch sphere $\hatb{n}\in \bS^2$ via the isomorphism between $\bC \bP^1$ and $\bS^3/U(1)$, i.e.,
\begin{align}
   & \Delta = |\Delta|e^{i\chi} Z  \\
   & Z =  \begin{bmatrix}
    \cos{\left(\dfrac{\theta}{2}\right)}e^{-i\phi/2} \\
    \sin{\left(\dfrac{\theta}{2}\right)}e^{+i\phi/2} \\
  \end{bmatrix} \nonumber \\
  &n_i = Z^\dagger \tau_i Z \nonumber
\end{align}
Here, defining $\chi_d$ and $\chi_g$ to be the phases of $D$ and $G$ respectively,  $\chi=\frac 1 2 [ \chi_d+\chi_g]$ and $\phi=[\chi_g-\chi_d]$ are the overall  and  relative SC phases and $\theta =2\arctan[|G|/|D|]$.
Note that OPs $\Delta$ which differ only by their global phases $\chi$ are mapped onto the same point on the Bloch sphere.

More generally, to the extent that it is possible to ignore variations in the magnitude of $\Delta$, the problem reduces to a  non-linear sigma model with a weakly broken $U(1) \times SO(3)$ symmetry, derived in Appendix \eqref{sec:nonlinear-sigma-derivation}:
\begin{align}
    \label{eq:nonlinear-sigma}
    F&= K_{\kappa} +K_{\kappa'} +K_{\hatb{n}}\\
    & +V_0 (|\Delta|)+\tilde{\bm{\alpha}} \cdot \hatb{n} + \frac{\tilde \beta_1}{2} [n_1]^2 + \frac{\tilde\beta_3}{2} [n_3]^2 +\ldots \nonumber\\
    K_{\kappa} &= \frac{\tilde\kappa}{2} \left|\vec \nabla \chi +\vec a- {\vec A}\right|^2 \nonumber \\
    K_{\kappa'} &= \tilde{\kappa}^\prime n_1 \prod_{\mu=x,y} \left[\partial_\mu \chi + a_\mu -\frac{[\hatb{n} \times \partial_\mu \hatb{n}]_1}{2 n_1}  -A_\mu \right] \nonumber \\
    K_{\hatb{n}} &= \frac{\tilde{\kappa}}{2} \left| \frac{\vec{\nabla}\hatb{n}}{2} \right|^2 \nonumber\\
    &+ \tilde{\kappa}'n_1  \left\{\frac{[\hatb{n} \times \partial_x \hatb{n}]_1}{2n_1} \frac{[\hatb{n} \times \partial_y \hatb{n}]_1}{2n_1} -\frac{\partial_x \hatb{n}}{2} \cdot \frac{\partial_y \hatb{n}}{2} \right\}\, , \nonumber
\end{align}
where $[\hatb{n} \times \partial_\mu \hatb{n}]_1=n_2 \partial_\mu n_3 -n_3 \partial_\mu n_2$ is the 1\ts{st} component of the vector $\hatb{n} \times \partial_\mu \hatb{n}$, while $\tilde \kappa\equiv |\Delta|^2\kappa$, $\tilde{\bm{\alpha}} \equiv |\Delta|^2\bra \alpha_1,0,\alpha_3+\beta_3'|\Delta|^2\ket$,  $\tilde \beta_j \equiv |\Delta|^4 \beta_j$, and $\ldots$  signifies terms that would come from higher order terms in the Ginzburg-Landau theory.
Importantly, $\vec a$ is the Berry connection associated with the motion of $\hat n$ on the Bloch sphere:
\begin{align}
    \vec a &\equiv Z^\dagger \left(\frac{1}{i} \vec \nabla Z \right)
\end{align}
The corresponding Berry curvature is related to the Pontryagin density
\begin{align}
    \vec \nabla \times \vec a = \frac{1}{2}
    \epsilon_{\mu\nu} \left[\hatb{n}\cdot (\partial_\mu \hatb{n} \times \partial_\nu \hatb{n}) \right]
\end{align}

\subsection{Ground State}
\label{sec:ground-state}
Let us first determine the ground state of a system in the presence of a uniform strain vector $\bm{\alpha}$ using the Ginzburg-Landau free energy \eqref{eq:GL-vectorized}.
For calculational convenience, we will consider the case $\beta_3'=0$ and $ \beta_1=\beta_3=\beta>0$ (in which case the free energy is invariant under a $U(1)$ symmetry associated with rotations of $\hatb{n}$ around $n_2$).
The general case is analyzed in Appendix \eqref{sec:general-ground-state}.

The potential term $V$ of the Ginzburg-Landau free energy can be rewritten as a sum of two terms,
\begin{align}
  V_2 &= -|\Delta|^2 (\alpha_0 +\bm{\alpha}\cdot \bm{m}) \nonumber \\
  V_4 &= \frac{1}{2}|\Delta|^4 \left(1 +\beta m^2 \right)
\end{align}
where $\bm{m}$ is the projection of the normalized vector $\hatb{n}$ onto the $\hatb{e}_1$-$\hatb{e}_3$ plane and $m\equiv |\bm m| =\sqrt{n_1^2 + n_3^2}$, where $n_j\equiv \hatb n\cdot \hatb{e}_j$.
The potential is minimized when $\bm{m}$ points in the same direction as $\bm{\alpha}$ so that the potential term is given by
\begin{equation}
  V = -|\Delta|^2 (\alpha_0 +\alpha m) +\frac{1}{2}|\Delta|^4 \left(1 +\beta m^2 \right)\ ,
\end{equation}
where $\alpha \equiv |\bm \alpha|$.  Since $0\leq m \leq 1$, the values of $|\Delta|$ and $m$ that minimize $V$ are uniquely determined: for $\alpha_0\geq \alpha/\beta \geq 0$,
\begin{equation}
  \label{eq:ground-state-1}
  |\Delta|^2 = \alpha_0, \quad m = \frac{\alpha}{\beta \alpha_0},
\end{equation}
which corresponds to two distinct values $\hat{\bm{n}}$  (related by time-reversal symmetry) with $n_2 = \pm\sqrt{1-m^2}$. For $\alpha/\beta > \alpha_0\geq -\alpha$, we have
\begin{equation}
  \label{eq:ground-state-2}
  |\Delta|^2 = \frac{\alpha_0 +\alpha}{1 +\beta}, \quad m=1,
\end{equation}
so that $n_2=0$.
Finally, $|\Delta|=0$ for $\alpha< -\alpha_0$.  Henceforth, we restrict our attention to the case $\alpha_0>0$ so $|\Delta|>0$ and that the nature of the ground state is determined by the value of $\alpha/\beta \alpha_0$.

In the limit of small strain $\alpha\ll \beta \alpha_0$ so that $m\sim 0$, $\hatb{n}$ points in the $\pm \hatb{e}_2$ direction, which corresponds to a TRSB $d\pm ig$ state.
Conversely, if $\alpha \ge \beta \alpha_0$ so that $m=1$, the Bloch vector $\hatb{n}$ points in the same direction as $\bm{\alpha}$ as denoted by the solid black dot in Fig. \eqref{fig:Bloch-domain-walls}.
This corresponds to a TRS preserving state determined by the the local strain.
From now on, when we discuss a situation in which the strain is nonzero, we shall implicitly assume that $\alpha \ge \beta \alpha_0$ and thus the uniform ground state is as in \eqref{eq:ground-state-2}.

\subsection{Domain walls}
\begin{figure}
\centering
\includegraphics[width=0.7\columnwidth]{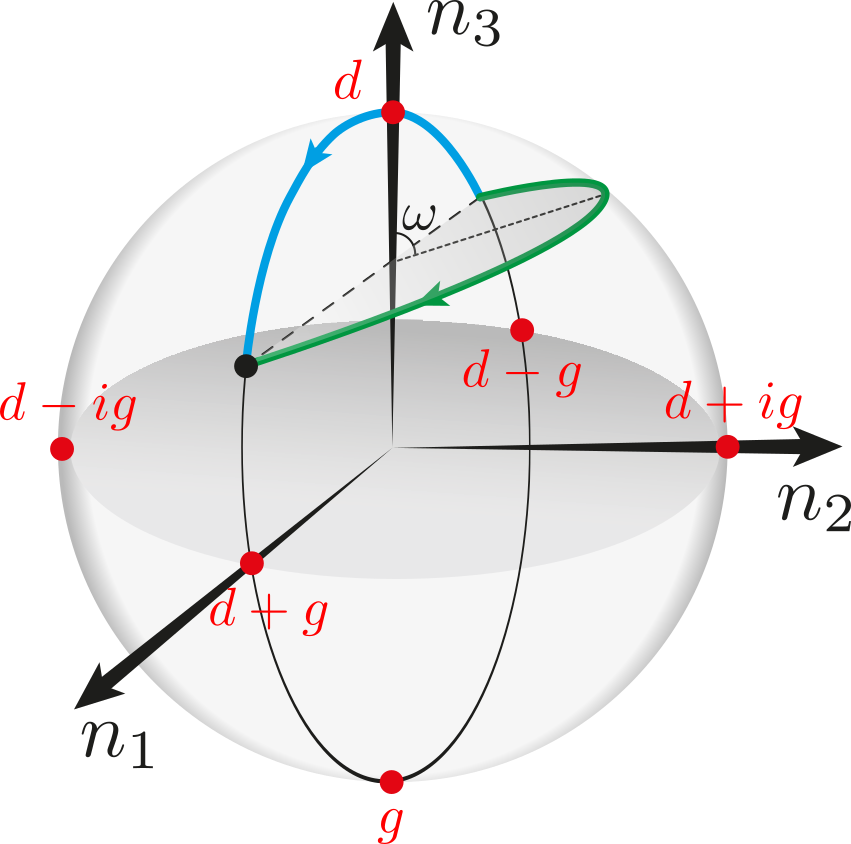}
\caption{Domain Walls. The figure is the Bloch sphere representation of possible transitions across the domain wall where $\alpha_1(y) \to \pm \ubar{\alpha}_1$ as $y\to \pm \infty$, and $\alpha_3(y) \equiv \ubar{\alpha}_3 >0$ is constant.
The blue arrow represents a TRS preserving transition through a pure $d$ state.
The green arrow represents a general TRSB transition restricted to a 2D plane intersecting the Bloch sphere, the angle of which relative to the $\hatb{e}_1$-$\hatb{e}_3$-plane is denoted by $\omega$.
In particular, if $\omega=\pi/2$, then only the relative phase $\phi$ changes from $\phi =\pi\to 0$.
}\label{fig:Bloch-domain-walls}
\end{figure}

We now consider the behavior of the order parameter $\Delta$ in the presence of spatially varying strain.  As it simplifies the analysis, we will do this in the context of the non-linear sigma model \eqref{eq:nonlinear-sigma}.
To begin with, we consider a domain wall separating  a region at $y <0$ in which the shear strain favors $d-g$ pairing ($\alpha_1 <0$) from a region at $y>0$ that favors $d+g$ ($\alpha_1 > 0$).
We consider the system to be translationally invariant in the directions parallel to the domain wall.

Far from  the domain wall, the relative phase $\phi(y)$ and amplitudes are determined by the strain.
Thus, at long distances, the only property of the order parameter texture that can depend on the nature of the domain wall is the change in the global phase, $\delta \chi$.
If we choose the global phase such that the order parameter is real at $y \to -\infty$, across the domain wall the order parameter must change from $d-g$ to $e^{i\delta\chi} (d+g)$.

If TRS is preserved everywhere (i.e. if the order parameter can be chosen to be real), then the only possible values of $\delta\chi$ are $0$ and $\pi$.
Any other value of $\delta\chi$ requires TRSB on the domain wall, which consequently means that there must be two symmetry related optimal values, $\pm \delta\chi$.

Below we discuss the derivation of $\delta\chi$ in several cases.  This is done by minimizing the free energy in Eq. \eqref{eq:nonlinear-sigma} in the presence of a given strain texture.
To be concrete, we will consider the case in which $\alpha_3(y) = \ubar \alpha_3$ is a constant and $\alpha_1(y)$ changes sign across the domain wall such that  $\alpha_1(y) \to \pm \ubar \alpha_1$ as $y \to \pm \infty$ with $\ubar \alpha > \beta \alpha_0$
\footnote{Note that the same sort of analysis can be applied to a domain wall across which $\alpha_3(y)$ changes sign with $\alpha_1=$ constant, e.g. between a region of $D+g$ pairing (indicating dominant $D$ and subdominant $g$) to $d+G$ pairing (dominant $G$).}.
The result can be expressed as a trajectory on the Bloch sphere, as shown in Fig. \eqref{fig:Bloch-domain-walls}, where the arrow indicates the direction of evolution  as $y$ varies from $-\infty$ to $+\infty$.

To begin with, consider some general results that follow without explicit calculation:
\begin{itemize}
  \item \textbf{Away from the multicritical point:}
  Consider the case in which $\ubar\alpha_3 > \beta \alpha_0$.
  In this case,  $|D|$ is everywhere larger than  $|G|$ and even at the ``center'' of the domain wall, defined as the point $y=0$ where $\alpha_1(0)=0$, there is no local tendency to TRSB.
  The optimal order parameter texture lies in the $\hatb e_1$-$\hatb e_3$ plane - as indicated by the blue trajectory in Fig. \eqref{fig:Bloch-domain-walls}.
  Here the $G$ component of the order parameter vanishes at $y=0$ and is negative on one side and positive on the other.
  Obviously, the analogous considerations apply for $\ubar\alpha_3 < -\beta \alpha_0$, with the role of $D$ and $G$ interchanged.
  In this case, TRS is preserved everywhere,  and $\delta \chi= \pi$.
  \item \textbf{Broad domain wall near the multi-critical point: }
  If $|\ubar\alpha_3| < \beta \alpha_0$, then near the center of the domain wall, the local terms in the free energy favor a TRSB solution, $d\pm ig$.
  Moreover, if the strain fields vary slowly on the scale of the superconducting coherence length, then the order parameter will be well approximated by the  uniform state corresponding to the local value of the strain.
  Thus,  the order parameter texture is not confined to the  $\hatb e_1$-$\hatb e_3$ plane, as shown by the green trajectory.
  This implies that both components of the order parameter remain non-zero everywhere, and thus that $\delta\phi = \delta \chi_g - \delta \chi_d = \pm \pi$.
  However, how much of this phase change is accommodated by changing the phase of $D$ or $G$ depends on energetics;  if the $D$ wave order is everywhere dominant, then $\delta \chi_d \approx 0$ and hence $\delta\chi \approx \pi/2$, while if the $D$ and $G$ are of nearly equal strength, then $|\delta \chi_g|\approx |\delta \chi_g|$ and hence $\delta\chi \approx 0$.
  Clearly, for intermediate cases, $0 < |\delta \chi|< \pi/2$.
  \item  \textbf{Narrow domain wall near the multicritical point:}
  Here, the nature of the solution  depends on a host of microscopic details.
  Since ``narrow'' and ``broad'' refer to the width of the domain wall relative to the superconducting coherence length, and given that the superconducting coherence length diverges as $T\to T_c$, at least near $T_c$ this is likely the most physically relevant situation.  We will thus treat this case more explicitly below.
\end{itemize}

\begin{figure}
\centering
\includegraphics[width=1\columnwidth]{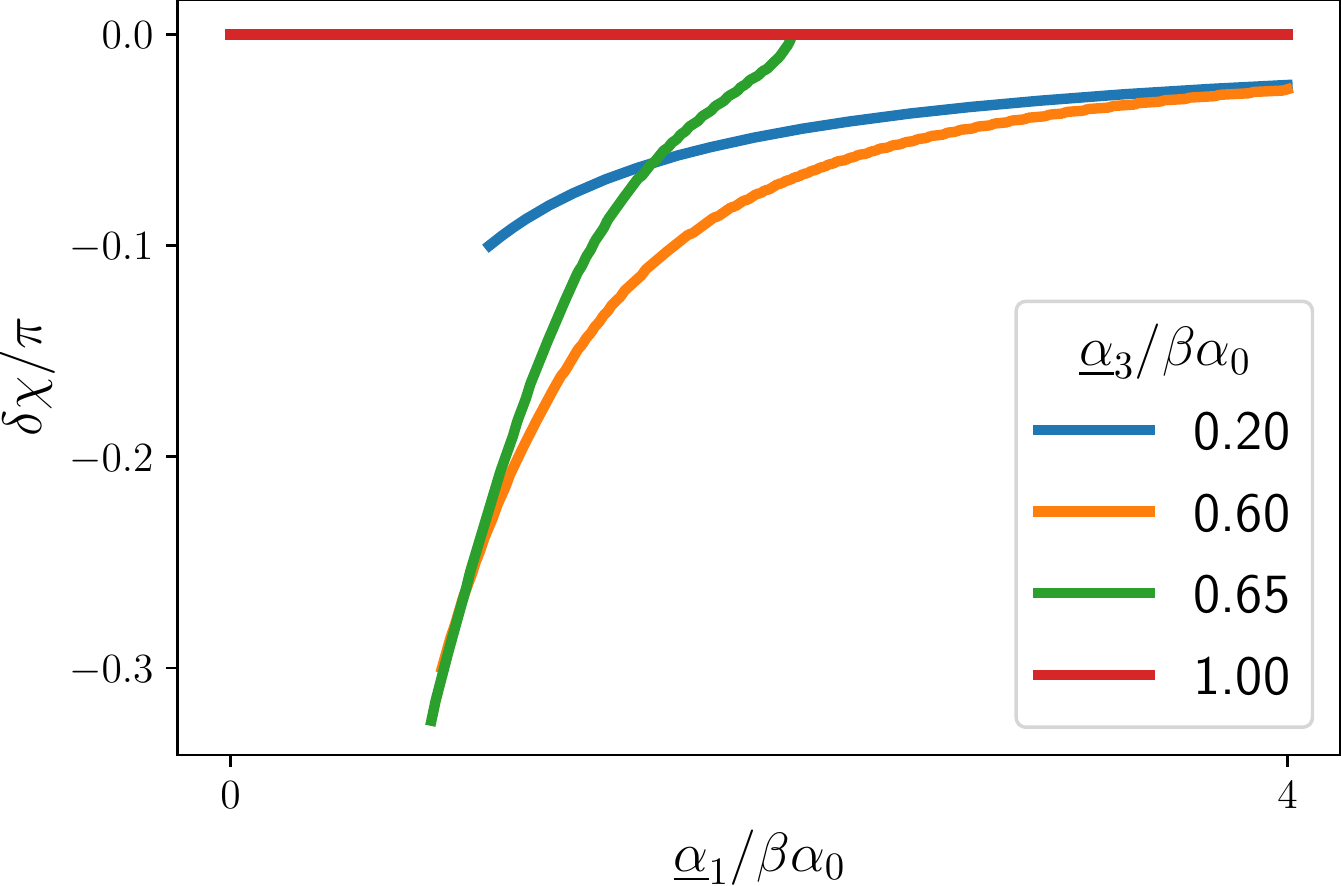}
\caption{Global phase. The change $\delta \chi$ in the global phase across a narrow domain wall where the $xy$ component of the strain, $\alpha_1$, is discontinuous. The domain wall is characterized by $\alpha_1 (y) = \ubar\alpha_1 \sign(y)$, while $\alpha_3(y) =\ubar\alpha_3$ is constant.
$\delta \chi$ is shown as a function of $\ubar{\alpha}_1/\beta \alpha_0$.
The different curves correspond to different fixed values of the uniform component of the strain, $\ubar\alpha_3/\beta \alpha_0$.
}
\label{fig:narrow-dchi}
\end{figure}

As an explicit model of a narrow domain wall, let $\alpha_1(\rvec ) = \ubar\alpha_1 \sign(y)$ where $\ubar \alpha_1>0$ and $\alpha_3(\rvec) \equiv \ubar{\alpha}_3 \geq 0$ be a constant.
We consider paths in which as $y$ goes from $-\infty$ to $\infty$, $\hatb{n}(y)$ follows a trajectory that lies in a plane $\omega$ intersecting the Bloch sphere - of the sorts illustrated by the different colored paths  in  Fig. \eqref{fig:Bloch-domain-walls}. 
When this plane is perpendicular to $\hatb{e}_2$, TRS is preserved (blue line) while all other trajectories break TRS.
A more complete solution of the problem does not result in qualitative changes in the conclusions.
With these simplifications, the domain wall energies $\Delta F$ can be computed analytically (see Appendix \eqref{sec:narrow-dw-calc}).

We can then find the plane $\omega$ with minimum domain wall energy for each set of values $\ubar{\alpha}_1, \ubar{\alpha}_3$ and compute the corresponding change in global phase $\delta \chi$, as shown in Fig. \eqref{fig:narrow-dchi}.
For $\ubar{\alpha}_3 \ge \beta \alpha_0$  (i.e., way from the multi-critical point), TRS is preserved everywhere, such that $\delta \chi = 0$.
Conversely, if $\ubar{\alpha}_3 <\beta \alpha_0$, (near the multi-critical point), TRS breaking near the domain wall is possible,
yielding $0<|\delta \chi| < \pi/2$.

\subsection{Topological Point Defects}

The domain walls we have discussed are natural strain patterns that are plausibly generic in real materials.  In addition, because there are two components of the strain-dependent vector, $\bm{\alpha}$, one can also conceive of 
vortex-like defects with  point-like cores  in 2d or line-like cores in 3d.  Here, we consider a pattern of strain such that along any path encircling the origin,  $\bm{\alpha}(\rvec)$ rotates by $2\pi$.  Given such a strain pattern, we can use the non-linear sigma model to explore the properties of the resulting SC order parameter texture that results.
\subsubsection{Vorticity and associated flux}

Far from the defect core, the form of the SC order parameter (up to its overall phase) is essentially determined by the local pattern of strain.
Regardless the SC order parameter texture, the current density,  $\vec J =- \delta F/\delta \vec A$, 
{can be computed exactly from the nonlinear sigma model \eqref{eq:nonlinear-sigma} since variations in magnitude $|\Delta|$ do not couple to the vector potential $\vec A$. 
}
From this it follows that
\begin{align}
  \label{eq:current}
  J_x &= \tilde \kappa [\partial_x \chi +a_x -A_x] \\
  &+\tilde{\kappa}^\prime n_1 \left[\partial_y \chi +a_y -\frac{[\hatb{n} \times \partial_y \hatb{n}]_1}{2n_1} -A_y \right] \nonumber
\end{align}
and similarly for $J_y$.
Far from the core under most circumstances, $\vec J \to \vec 0$; indeed, beyond a London penetration depth it vanishes exponentially.  Thus, we can invert Eq. \eqref{eq:current} to obtain an expression for $\vec A$ in terms of the SC order parameter texture valid wherever $\vec J$ is negligible.
Then, by integrating the vector potential $\vec{A}$ along a contour $C$ that encloses the origin at a distance, we obtain an expression (modulo an additive integer) for the enclosed flux $\Phi$
in units of the superconducting flux quantum $\Phi_0=h/2e$:
\begin{align}
  \label{eq:flux}
  \frac{\Phi}
  {\Phi_0}& = \frac{\Omega}{4\pi} - \frac{\kappa'}{4\pi} \oint_{C} [\hatb{n} \times \partial_\mu  \hatb{n}]_1 T_{\mu\nu}\ dr_\nu   \\
  T &=
  \begin{bmatrix}
    \kappa^\prime n_1 & \kappa \\
    \kappa & \kappa^\prime n_1
  \end{bmatrix}^{-1} \nonumber
\end{align}
where $\Omega$ is the solid angle enclosed by the contour of $\hatb{n}$ on the Bloch sphere.

Note that when $[\hatb{n} \times \partial_\mu  \hatb{n}]_1=0$ along $C$, the second term vanishes and thus the flux quantum captured is expressed
entirely in terms of the Berry phase $\Omega/4\pi$.  In particular, since strain stabilizes a TRS preserving state, we will typically be interested in situations in which $\hatb{n} \cdot \hat{\bm{e}}_2=0$ far from the defect core, insuring that this condition is satisfied.  In this case, whenever $\hatb{n}$ follows a  trajectory that encircles the origin (as in Fig. \eqref{fig:GL-vortex-b}), there must be an associated half-flux quantum of flux bound to the defect.

\subsubsection{Example of a strain-induced half-quantum fluxoid}
\begin{figure}
  \centering
  \centering
  \subfloat[\label{fig:GL-vortex-a}]{%
    \centering
    \includegraphics[width=.6\columnwidth]{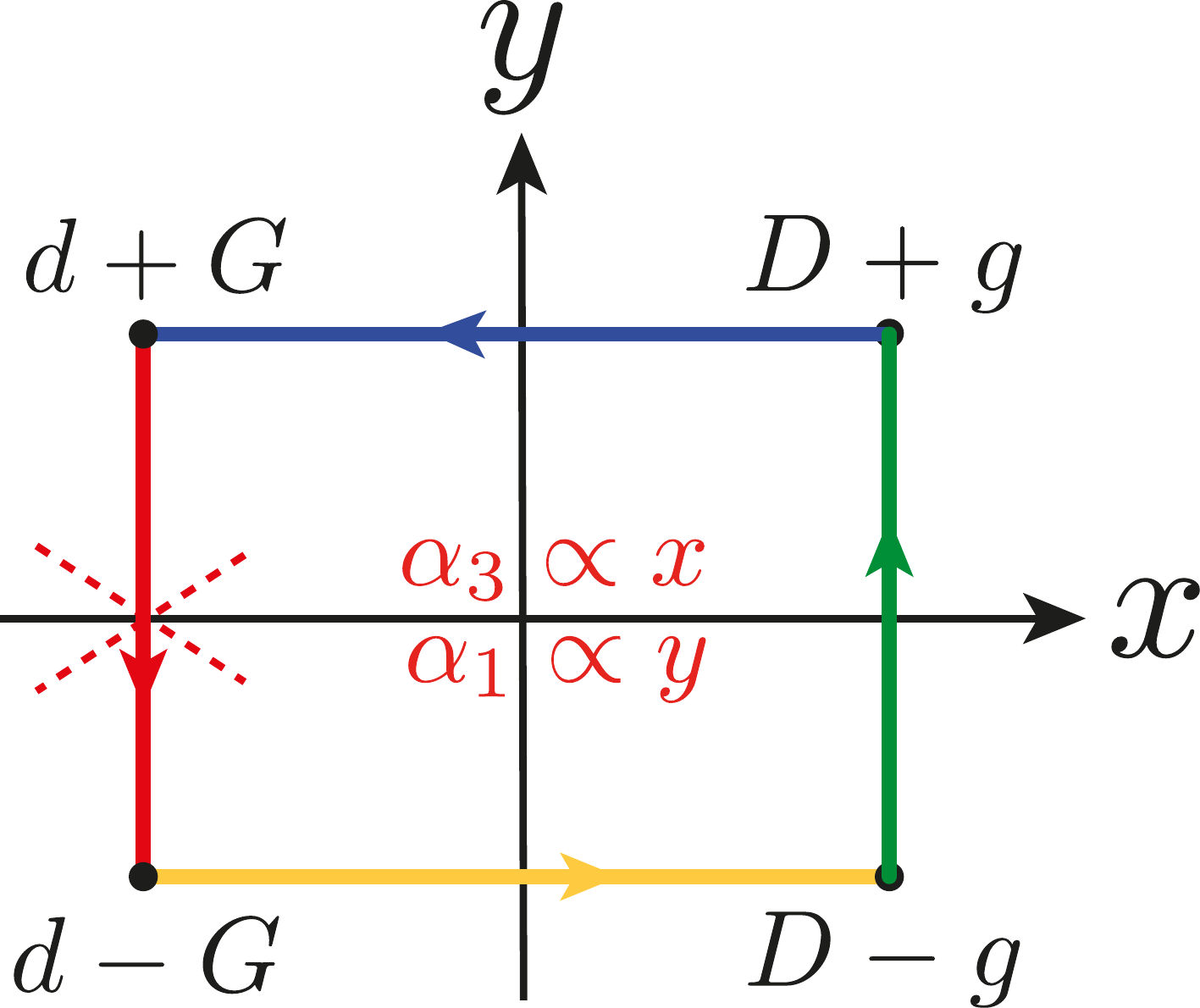}
  }\hfill
  \subfloat[\label{fig:GL-vortex-b}]{%
    \centering
    \includegraphics[width=.7\columnwidth]{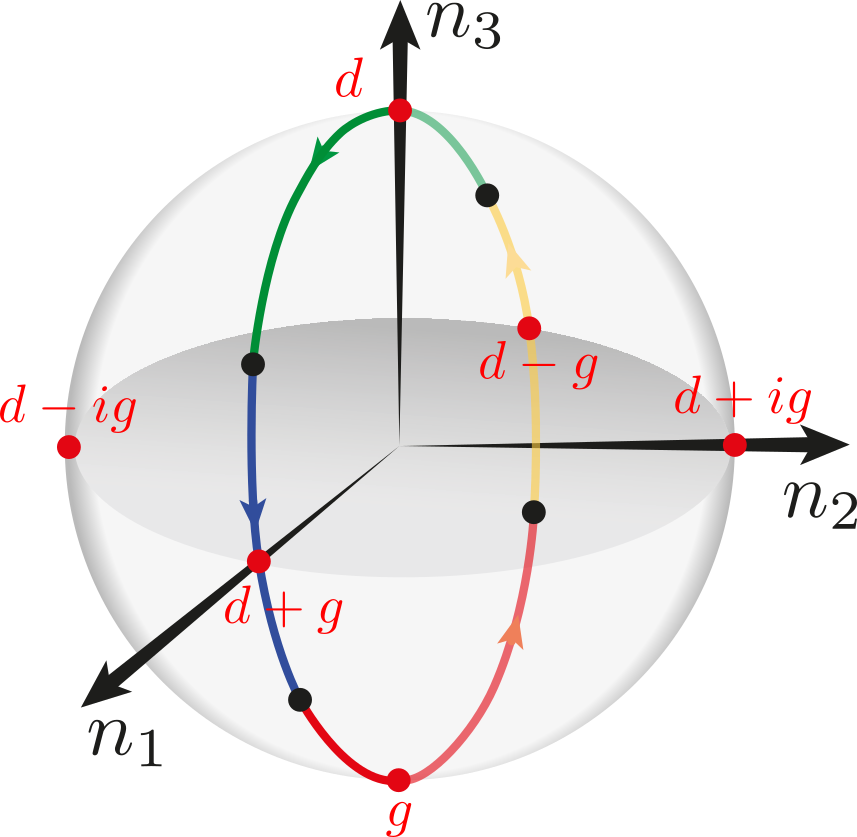}
  }\hfill
  \caption{A strain induced half quantum fluxoid.
  (a) Real space contour around a topological point defect, along which the strain vector $\bm{\alpha}$ winds by $2\pi$. The dashed cross shows the location of a $\pi$ mismatch in the phase of the order parameter. $D+g$ in the $x>0$, $y>0$ quadrant represents order parameter with a dominant $D$ component and a smaller $G$ component, and similarly in the other quadrants.
  (b) Corresponding trajectory of the order superconducting order parameter along the Bloch sphere. TRS is preserved along the path.
  Each segment of the contour is color coded so that the same colors in the bottom and top figure correspond to each other.
  }\label{fig:GL-vortex}
\end{figure}

We now consider an explicit version of such a strain texture (Fig. \eqref{fig:GL-vortex-a}), consisting of four domains separated by domain walls that intersect at the origin.
Thus, we take $\alpha_3(x,y)= \ubar\alpha_3 F_3(x)$ and $\alpha_1(x,y) = \ubar \alpha_1 F_1(y)$, where $F_j(r) = -F_j(-r)$, and $F_j(r) \to 1$ as $r\to \infty$.
We further assume that $\ubar \alpha_j > \beta \alpha_0$ -- i.e. far away from the origin we are in the large strain regime where the preferred order parameter is set by the strain and TRS is preserved.
As indicated by the labels, the $D$ component is enhanced compared the $G$ component for $x>0$. For  $x<0$, the $G$ component is favored. The combination $d+g$ is favored for $y>0$, while $d-g$ is favored for $y< 0$.

Let us now consider a closed path $C$ encircling the origin (Fig. \eqref{fig:GL-vortex-a}), where the preferred SC state (up to the global phase) is denoted in each quadrant, e.g., $D+g$ denotes a TRS preserving SC state with dominant $D$ component.
Since the contour $C$ is far away from the origin, the domain walls between quadrants are narrow,
and the strain is always sufficiently large such that TRS breaking is never favored locally.
Fig. \eqref{fig:GL-vortex-b} shows the trajectory of the order parameter on the Bloch sphere, in which each segment of the contour is color coded to correspond to that in the top diagram.
We then see that the OP $\Delta$ wraps around by $2\pi$ while being confined to the $\hatb{e}_1$-$\hatb{e}_3$ plane and thus $[\hatb{n}\times \partial_\mu \hatb{n}]_1 =0$ along the contour.
Eq. \eqref{eq:flux} thus implies that this strain texture captures a half quantum of flux.

At an intuitive level, the same conclusion can be reached by considering the nature of the order parameter texture along the various line segments in Fig. \eqref{fig:GL-vortex-a}. Since it is energetically favorable to keep the dominant piece of the SC order parameter uniform, the overall phase ($\delta\chi$) is constant along any of the segments other than the red one, along which the dominant portion of the order parameter changes sign, favoring $\delta\chi = \pi$.  Of course, this change in phase will in actuality be spread out along the entire path, but this argument captures the $\pi$ phase mismatch along the close path that results in a half quantum vortex.

\section{Microscopic  Analysis}
\begin{figure}[ht]
\subfloat[\label{fig:phase-diag-a}]{%
  \centering
  \includegraphics[width=.8\columnwidth]{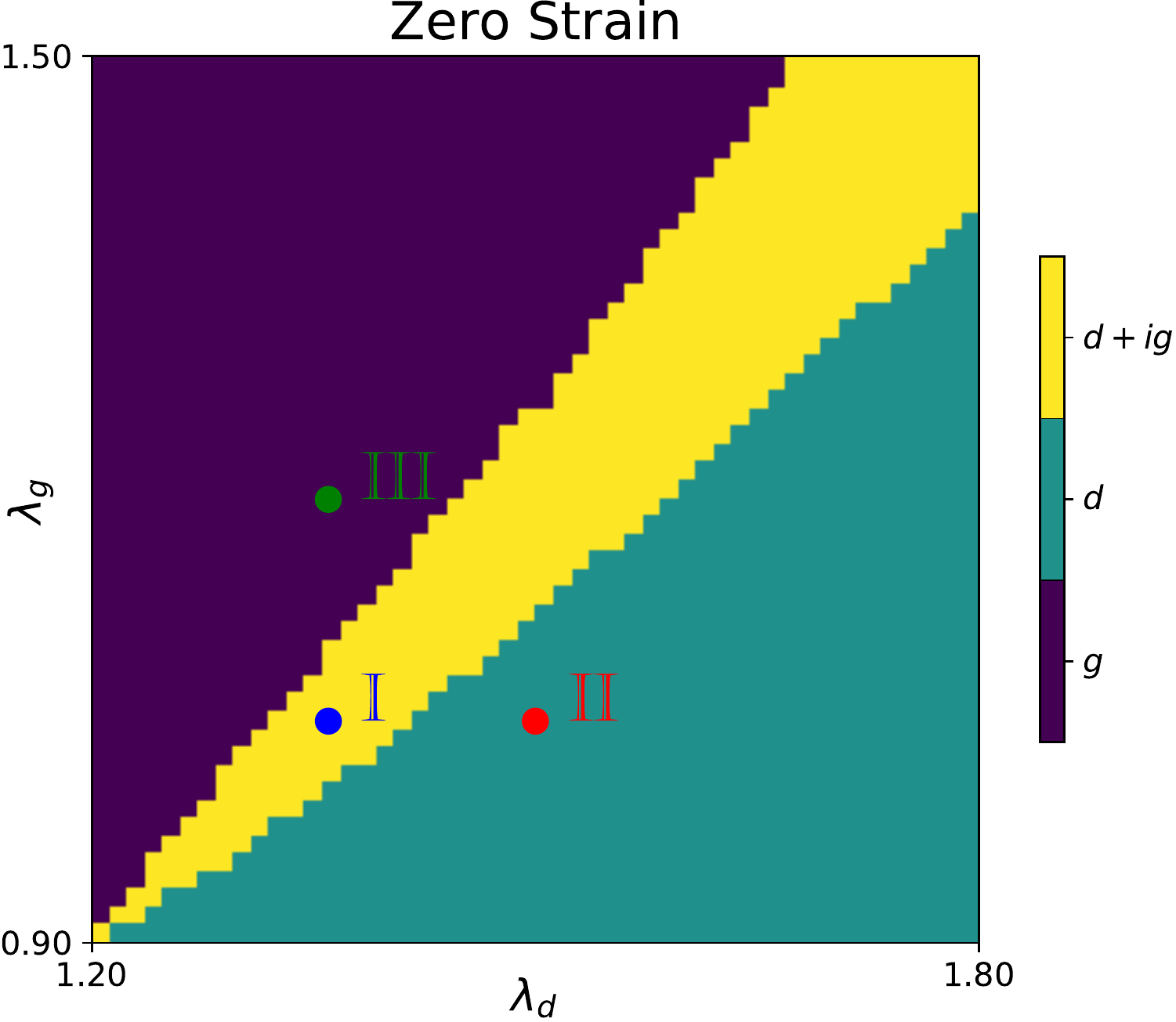}
}\hfill
\subfloat[\label{fig:phase-diag-b}]{%
  \centering
  \includegraphics[width=1\columnwidth]{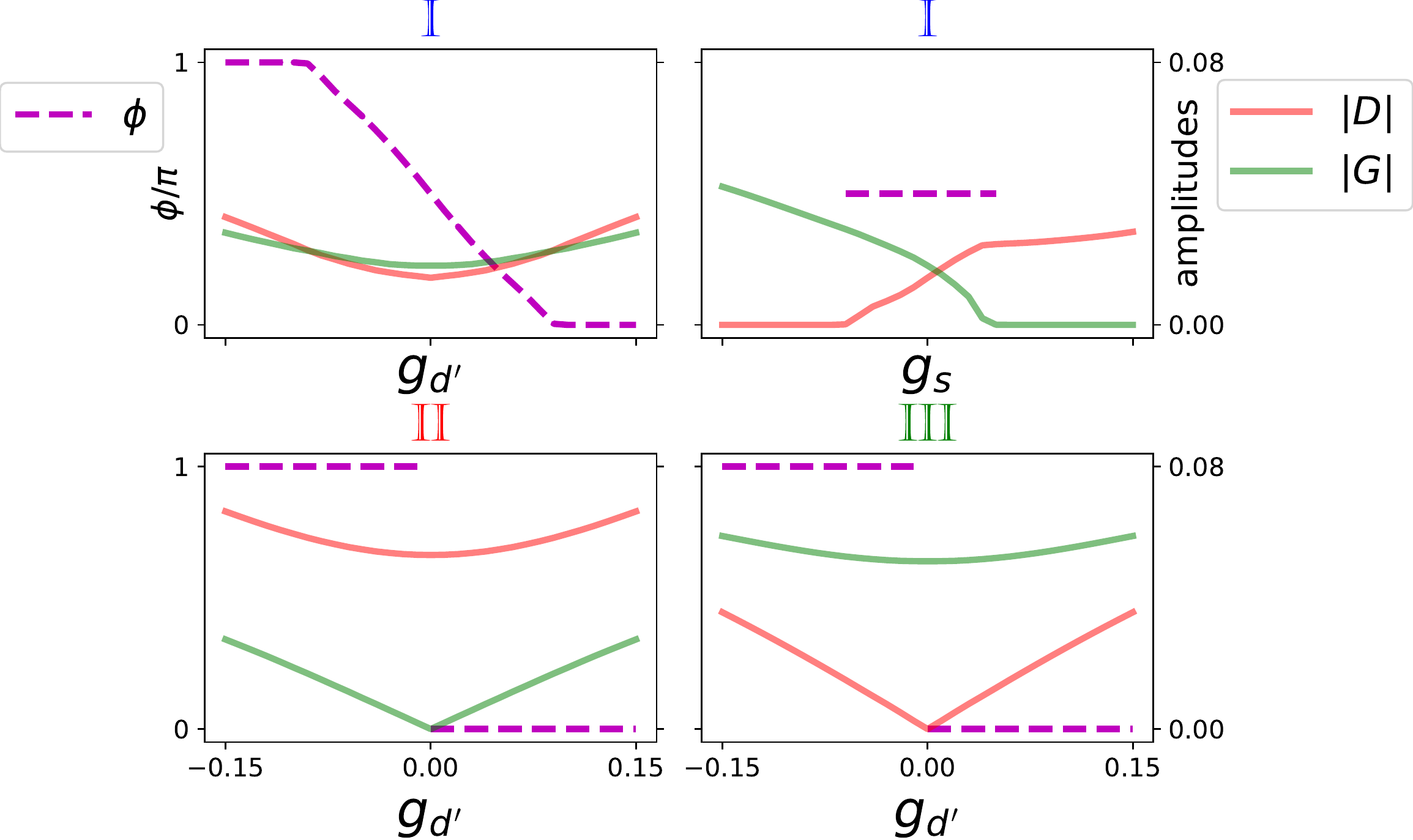}
}\hfill
\caption{
(a) Phase diagram of the microscopic Hamiltonian, Eq.~(\ref{eq:H0},\ref{eq:H1}), as a function of the interaction strengths $\lambda_d,\lambda_g$, calculated within self-consistent mean-field theory, in the absence of strain.
Near degeneracy $\lambda_d \sim \lambda_g$, the yellow portion denotes the TRSB $d+ig$ state, while the cyan and purple regions represents a pure $d$ or pure $g$ state, respectively.
3 representative point (I, II, III), one in each phase, are used in further calculations.
(b)
The top two panels depict the evolution of the uniform state at point (I) with shear $g_{d'}$ and isotropic $g_s$ strain.
The dashed line represents the relative phase $\phi$ given by the left $y$-axis, while the solid (red, green) lines represent the amplitudes $|D|,|G|$, given by the right $y$-axis.
Similarly, the bottom two panels represent the evolution of the uniform state for sample points (II) and (III) under shear $g_{d'}$ strain.
}
\label{fig:phase-diag}
\end{figure}

We now address these same issues from a more microscopic perspective in the context of Bardeen-Cooper-Schrieffer (BCS) mean-field theory.  Specifically, we solve the self-consistency equations for a generic 2D Hamiltonian with attractive 
$d$ and $g$-wave interactions, and with spatially varying band-structure parameters that encode the same patterns of spatially varying strain discussed above.

Let the full Hamiltonian $H_\text{full} = H_0 +H_1$ be defined on a 2D square lattice.
The free Hamiltonian $H_0$ is characterized by the (single-band) TRS-preserving hopping matrix $t$ so that
\begin{equation}
\label{eq:H0}
H_0 = -\sum_{\rvec',\rvec,s} \left[t(\rvec',\rvec) + \mu\right] c_{\rvec' s}^\dagger c_{\rvec s}^{\vphantom{\dagger} }
\end{equation}
and the attractive (pairing) interaction term $H_1$, assumed to be spatially uniform, is
\begin{align}
\label{eq:H1}
H_1 &= -\sum_{\tau = d,g}\lambda_\tau \sum_{\rvec} P^\dagger_\tau(\rvec) P_\tau^{\vphantom{\dagger} } (\rvec) \\
P^\dagger_\tau(\rvec) &= \sum_{\rvec',s,s'} f_\tau(\rvec'-\rvec) c^\dagger_{\rvec' s'}[i\sigma_2]_{s's}c^\dagger_{\rvec s} \nonumber
\end{align}
where $\lambda_\tau >0$ with $\tau=d$ or $g$ encodes the strength of the interaction in the designated symmetry channels.  Here $f_\tau$ are TRS preserving form factors  that transform according to the requisite distinct irreducible representations of the point-group symmetry:
\begin{align}
   f_d (\rvec) &= \frac 14 \delta(r=1) \left[x^2-y^2\right] \\
   f_g (\rvec) &= \frac{3\sqrt{3}}{32}\delta(r=\sqrt{5}) [xy (x^2-y^2)] \nonumber
\end{align}
where the factor of the Kronicker-$\delta$ in each expression is 1 when $\rvec$ connects, respectively, first 
and fourth nearest-neighbor sites.  The hopping matrix elements can likewise be expressed in terms of these and the local strain as
\begin{align}
    t(\rvec+\rvec',\rvec)&= \delta(r'=1) + t\delta(r'=\sqrt{2})\\
    & +g_s(\rvec) \odot f_s(\rvec')+ g_{d'}(\rvec) \odot f_{d'}(\rvec')\nonumber
\end{align}
where $g_s(\rvec)$ and $g_{d'}(\rvec)$ parameterize, respectively, the spatial profile of the isotropic and shear strain, and {$\odot$ is the symmetrization of the product term, e.g., $g_s(\rvec) \odot f_s (\rvec') \equiv \frac{1}{2} [g_s(\rvec) +g_s(\rvec+\rvec')]f_s(\rvec')$. }
Here, $f_s(\rvec)$ and $f_{d'}(\rvec)$ are form factors with isotropic and shear $[xy]$ symmetry of the underlying lattice, which we take to be
\begin{align}
     f_{s}(\rvec) = \delta(r=2), \quad f_{d'}(\rvec) = \delta(r=\sqrt{2}) \left[xy\right]
\end{align}

We  construct a  mean-field
BCS trial Hamiltonian $H$, given by 
\begin{align}
H &= \sum_{\rvec',\rvec,s} \cT(\rvec,\rvec) c_{\rvec s}^\dagger  c_{\rvec s}^{\vphantom{\dagger} } \\
&+ \sum_{\rvec',\vec r} \left[\Delta(\rvec',\rvec ) c_{\rvec'\uparrow}^\dagger  c_{\rvec\downarrow}^\dagger + \text{h.c.}\right] \nonumber
\end{align}
The full self-consistency field equations (SCFs) can then be derived by extremizing the resulting variational free energy in the standard fasion -- details  are given in Appendix \eqref{sec:BCS-calc}.
It should be noted that in order to guarantee that the results satisfy the equation of continuity,  i.e., $\vec\nabla \cdot \vec J=0$, it is generally insufficient to only solve the SCFs for the gap function - both $\cT$ and $\Delta$ must be determined self-consistently  (see Appendix \eqref{sec:zero-div-proof} for a proof).

\subsection{Uniform states}
\label{sec:uniform-states}

To begin with, we study the uniform case tuned close to the multi-critical point.
In Fig. \eqref{fig:phase-diag-a} we show the  mean-field ground-state phase diagram of the microscopic model defined above as a function of the pairing interactions, $\lambda_d$ and $\lambda_d$, in the absence of ``strain'' (i.e. for $g_s=g_{d'}=0$), for $t=0.4$ and for the chemical $\mu$ chosen so that the mean electron density per site is $n\approx 0.3$.
There are three distinct phases in this case:  a pure $d$ wave phase for $\lambda_d $ sufficiently larger than $\lambda_g$, a pure $g$ wave phase for sufficiently large $\lambda_d$, and a relatively narrow coexistence phase centered at the the line $\lambda_d=\lambda_g$. The latter phase has a relative phase $\phi=\pm \pi/2$, i.e., it is a $d\pm ig$ phase, for all parameters studied here.

To illustrate the effect of shear strain, we chose representative points in the phase diagram indicated by the three points in Fig. \eqref{fig:phase-diag-a}, and explore the evolution of the ground-state order upon application of uniform strain, i.e. non-zero $g_{d'}$ or $g_s$.
Shown 
in Fig. \eqref{fig:phase-diag-b} are the magnitude of the $d$ and $g$ components of the order parameter, $|D|$ and $|G|$, as well as the relative phase, $\phi$, for these three cases:

\begin{itemize}
  \item
  The top two panels of Fig. \eqref{fig:phase-diag-b} show the evolution with strain of the case in which we are most interested - the strain-free ground state has $d\pm ig$ pairing.
  As the shear strain, $g_{d'}$, is varied,  $|D|$ and $|G|$ remain comparable, although both increase slightly, roughly in proportion to $|g_{d'}|^2$ - which is a density of states effect.
  More dramatically, the relative phase evolves smoothly, up to a critical value at which TRS is restored, i.e. where $\phi$ reaches either 0 or $\pi$,  which marks the point of a transition to $d+g$ or $d-g$ pairing respectively.
  In contrast, as a function of the  isotropic strain $g_s$, the evolution from the $d+ig$ state to a pure $g$ or pure $d$ state involves a change of the relative amplitudes $|D|$ and $|G|$,  while the relative phase $\phi = \pm \pi/2$ remains constant.
  \item
  The lower two  of Fig. \eqref{fig:phase-diag-b} represent the shear strain evolution under conditions in which  at zero strain either the $d$ or $g$  component is absent.  In both cases, the component that is dominant at zero strain remains dominant;  indeed, its overall magnitude increases in much the same way as in the top panel.
  As required by symmetry, the component that vanished in the absence of strain exhibits an initial linear increase in magnitude with increasing strain.
  However, in this case, the relative phase is a discontinuous function of strain;  the ground-state always preserves TRS and jumps from being $d+g$ to $d-g$ as the sign of the strain changes.
\end{itemize}

We relate these results to the Ginzburg-Landau theory as follows: In the absence of strain, the system is tuned to the tri-critical condition $\alpha_1\approx\alpha_3\approx 0$ when $\lambda_d\approx \lambda_g$.
Moreover, since the system exhibits  $d+ig$ order in this case, the requisite inequalities $\beta_1>0$, $\beta_3 >0$ are automatically satisfied.  The shear strain can be identified with $\alpha_1 \propto g_{d'}$.
Conversely, even relatively small values of $|\lambda_d-\lambda_g|/|\lambda_d+\lambda_g|> 0.1$ are enough to produce a sufficiently large value of $\alpha_3$ such that even 
at $T=0$, $|\alpha|> \beta/|\alpha_0| $.
Also notice that we have taken relatively strong interactions; this condition becomes exponentially more restrictive the weaker the overall coupling, $|\lambda_d+\lambda_g|$.
The extent to which the system needs to be ``fine-tuned'' near to the tetra-critical point is quantified as the narrowness of this coexistence phase in the zero-temperature phase diagram.

\subsection{Domain Walls}
\begin{figure}
\centering
\includegraphics[width=1\columnwidth]{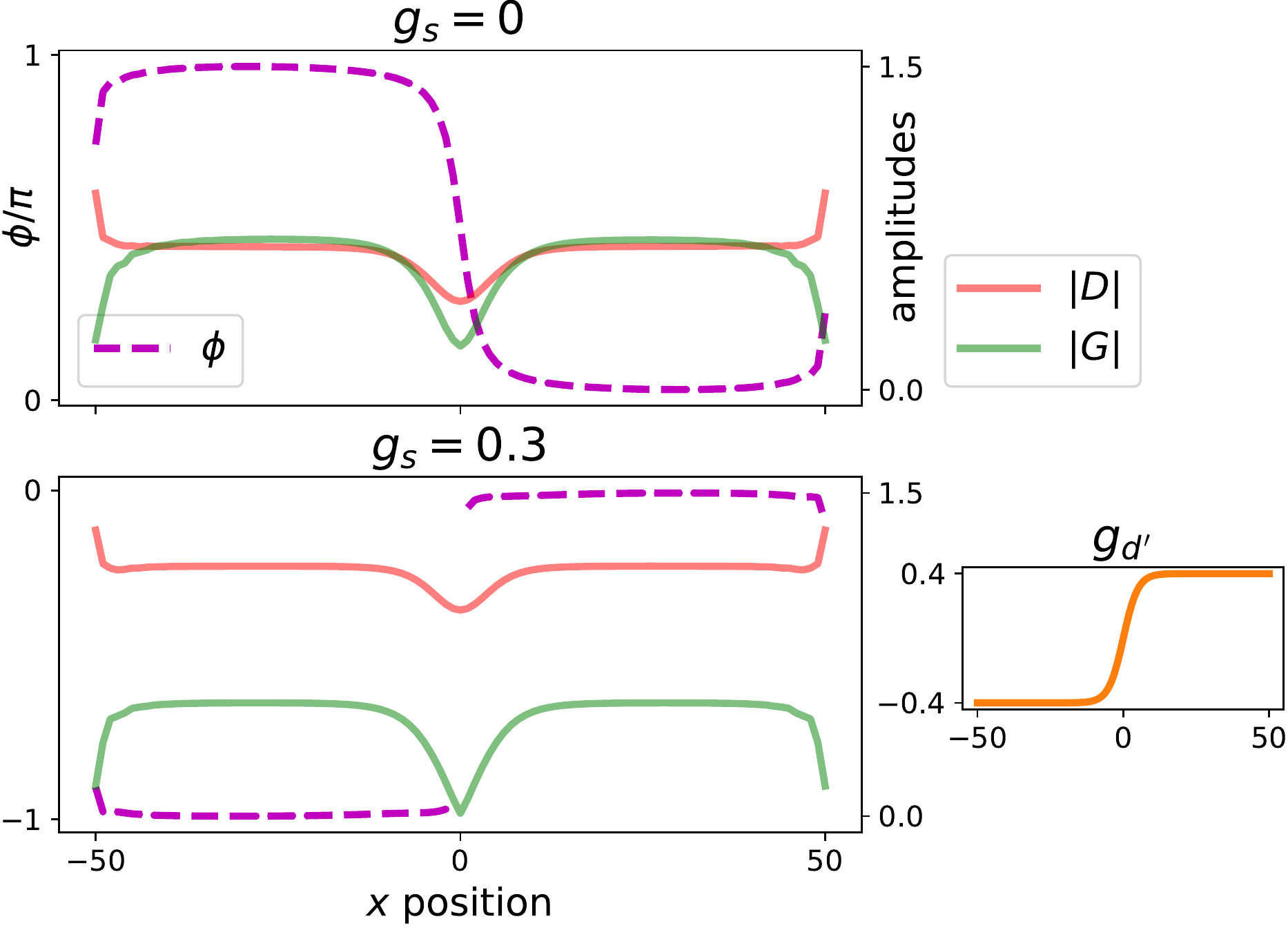}
\caption{Domain Walls. The top panel shows the 
behavior of the SC order parameter across a domain wall at which the shear strain $g_{d'}(x)$ changes sign when the system is detuned  
from the multi-critical point 
such that $g_s(x)=0.3$.
The bottom panel 
shows the same sort of domain wall for a system near the multi-critical point 
so that $g_s(x)=0$.
The spatial variation of $g_{d'}$, 
shown in the mini-plot,
is of the form $g_{d'}(x)=0.4 \tanh(x/l)$ where $l=5$ and $|x| \le L =50$.
The dashed lines represent the relative phase $\phi$ with values given by the left $y$-axis, while the solid lines are the amplitudes of the $D$ and $G$ components with values given by the right $y$-axis.
}
\label{fig:domain-wall}
\end{figure}
We next 
address the domain wall behavior for a system near and away from the multi-critical point.
To circumvent the difficulty of large coherence lengths at small interaction strength, we use 
large interaction strengths $\lambda_d = 3.2$ and $\lambda_g =4.9$, 
chosen such that in the absence of strain, the uniform system is near the multi-critical point, i.e.., in a $d+ig$ state.
We then introduce an $x$-dependent shear strain with a domain wall along the $y$-axis, along which the system is translationally invariant.
As a specific example, we take $g_{d'}=0.4 \tanh(x/l)$ where $l=5$ and $|x|\le L=50$.  We take $g_s(x)$ to be constant.

Fig. \eqref{fig:domain-wall} (top) shows the profiles of the order parameters and relative phase $\phi$ as a function of $x$
in the case $g_s = 0$, i.e., near the multi-critical point. 
$\phi$ twists through $\pi/2$ passing through the domain wall, indicating the TRS is broken there. 
Fig. \eqref{fig:domain-wall} (bottom) shows a TRS preserving 
domain wall for $g_s = 0.3$, i.e., away from the multi-critical point.

\subsection{Half quantum vortex}
\begin{figure}
\subfloat[\label{fig:BCS-vortex-a}]{%
  \centering
  \includegraphics[width=1\columnwidth]{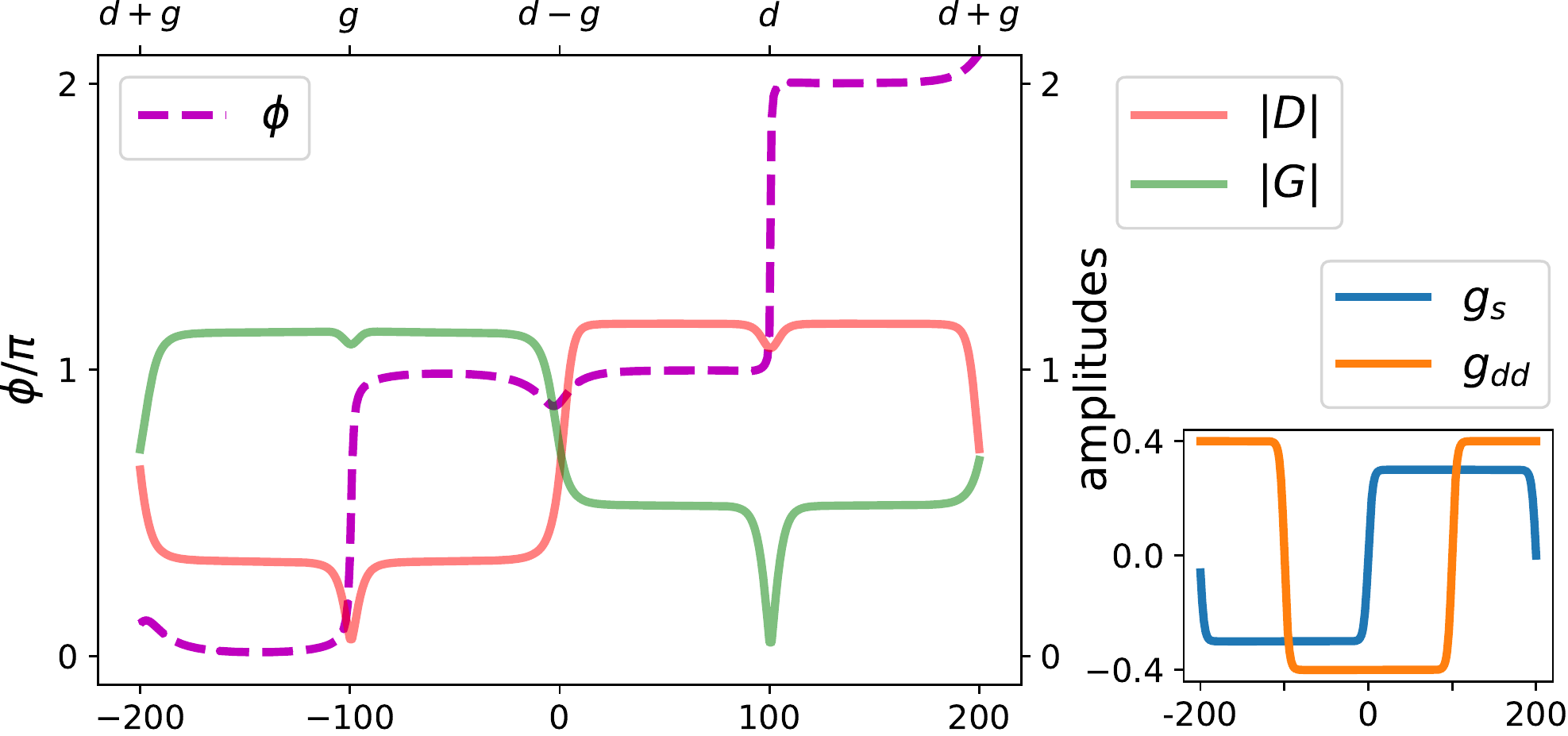}
}\hfill
\subfloat[\label{fig:BCS-vortex-b}]{%
  \centering
  \includegraphics[width=0.8\columnwidth]{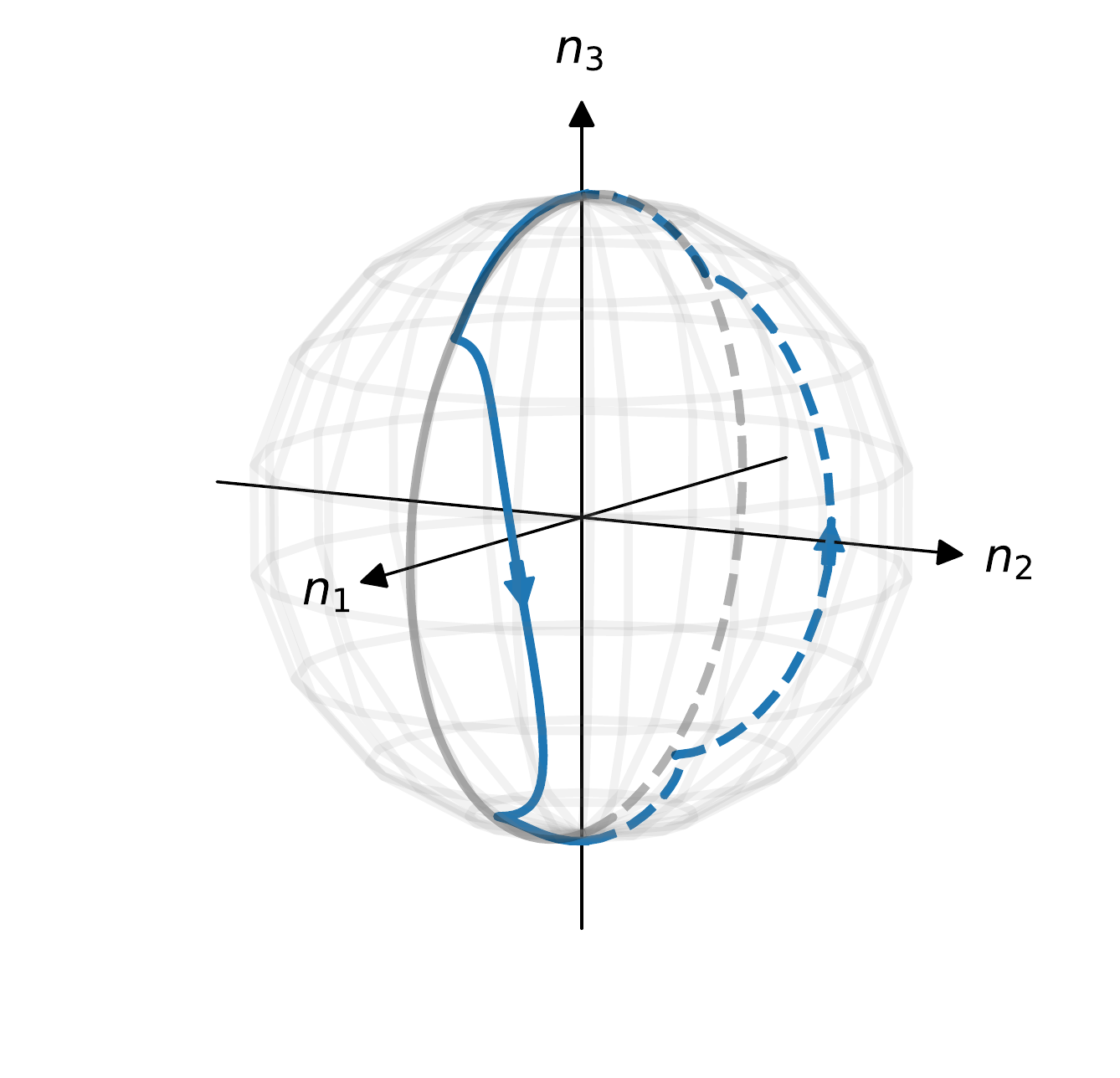}
}\hfill
\caption{Half quantum vortex pinned by strain texture.
(a) The behavior of the OP $\Delta =(D,G)$ as a function of $x$.
The two components of the strain, $g_s(x)\propto \alpha_3$ and $g_{d'}(x)\propto \alpha_1$, are shown on the right. The domain walls are of the form $\pm \tanh(x/l)$ with $l=5$, and the cylinder's circumference is $400$ sites.
The dashed line represents the relative phase $\phi$ given by the left vertical axis, while the solid lines represent that amplitudes $|D|$ and $|G|$ (right vertical axis).
The bottom horitzontal axis represents the $x$ position, while the top horizontal axis is labeled with the uniform state corresponding to the given strain texture at that position.
(b). Trajectory of the OP $\Delta=(D,G)$ projected onto the unit Bloch sphere (blue line). The arrow represent the direction of the contour. The gray circle line emphasizes the $\hatb{e}_1$-$\hatb{e}_3$ plane.
}
\label{fig:BCS-vortex}
\end{figure}

Finally, to construct a tractable situation in which the previously discussed phenomenological analysis leads to a half-quantum vortex, we consider the previous system on an infinite cylinder
with periodic boundary conditions in the $x$-direction and translational invariance along the $y$-direction.
The circumference of the cylinder (in the $x$-direction) is 400 sites.
We then vary the microscopic terms [$g_s(x)$ and $g_{d'}(x)$] corresponding the evolution of $\alpha_1$ and $\alpha_3$ shown in Fig. \eqref{fig:GL-vortex}.
Solving the full SCF equations, we obtain $D(x)$ and $G(x)$ as shown in Fig. \eqref{fig:BCS-vortex}.

The results corroborate the expected behavior from the Ginzburg-Landau theory.  Notice that the $x$-labels on top of Fig. \eqref{fig:BCS-vortex-a} indicate the expected state at each position along the contour.
Most importantly, the relative phase $\phi$ winds by $2\pi$ around the cylinder, from which it follows that $\delta \chi=\pi$. This result is translated into a trajectory on the Bloch sphere in Fig. \eqref{fig:BCS-vortex-b}. Note that the trajectory of the order parameter is close to the $\hatb{e}_1$-$\hatb{e}_3$ plane, but deviates from it somewhat in parts of the trajectory, indicating that TRS is broken in certain regions along the path. Correspondingly, in this calculation, we expect the flux induced by the strain pattern to deviate slightly from $\Phi_0/2$ [see Eq.~\eqref{eq:flux} and the discussion that follows].

\section{Conclusions}

In this paper we have primarily explored the effect of inhomogeneous strain on the superconducting OP textures of a system tuned close to a tetra-critical point at which two different superconducting components have equal $T_c$'s.  Such a tetracritical point can arise due to symmetry - when in the absence of strain the two components form a two dimensional irreducible representation of the point group symmetries.  However, we have primarily focused on the case in which the tetracritical point arises from tuning a symmetry preserving parameter near to a critical value - a situation that might arise accidentally in a small subset of superconducting materials.

In this situation, even relatively small local strain can readily detune the system from the tetracritical condition sufficiently that  a single component OP is locally favored.  However, if on average the system is near enough to this critical value, it is also natural to find domain walls between regions of different dominant strain in which the system is approximately tetra-critical.  This, in turn, can lead to TRS breaking on such domain walls, and thus to a state which globally breaks TRS but in which the symmetry breaking is locally significant only on a network of such domain walls.  We have also shown that appropriate patterns of inhomogeneous strain can bind a half-quantum vortex.

These results may have interesting implications for a number of materials that show evidence for either an exact or a near-degeneracy of two SC orders, including UPt$_3$ \cite{jin1992uniaxial, luke1993muon, schemm2014observation}, URu$_2$Si$_2$ \cite{yamashita2015colossal,schemm2015evidence}, UTe$_2$ \cite{ute2}, doped Bi$_2$Se$_3$ \cite{matano2016spin,Fu2014}, certain Fe based superconductors \cite{fernandes2014drives, grinenko2020superconductivity, bohm2014balancing}, and of course SRO. Most importantly, such a near degeneracy necessarily implies an enhanced sensitivity to variations in local strain, which can lead to a variety of otherwise unexpected behaviors. 
Note, however, that half quantum vortices and other topological defects can also arise as dynamical excitations  in 
multi-component superconductors in the absence of any strain effects~\cite{volovik1976line,Sigrist1989,Chung2007,Wu2017,Etter2020}.

It is worth noting that while our results are quite general, they are obtained within mean-field theory, and neglect thermal fluctuations of the superconducting order parameter. These can lead to interesting effects close to the $T_c$ in multi-component superconductors~\cite{Fischer2016,Yu_2019,hecker2018vestigial,Cho2019}.

This study was undertaken with the SC state of SRO in mind.  It is well established that the SC state is highly strain-sensitive.  There are also a variety of experimental observations - ultra-sound anomalies key among them - that are most naturally consistent with an assumed near degeneracy between a $d$ and $g$ wave SC component.
However, a variety of other experimental results appear, at first, difficult to reconcile with this scenario \cite{newhicks,nelson2004odd,kidwingira2006dynamical,jang2011observation}.  The present results 
suggest a route to understanding some of these additional observations.  This includes a suppressed thermodynamic signature of the TRS breaking transition 
and the possibility of half quantum vortices, even though some aspects of the actual experiment~\cite{jang2011observation} - for instance the dependence on an in-plane field - still need be addressed.

As mentioned in the introduction, a key issue concerns the strain-induced splitting between the SC and the TRS breaking transitions.
It has been found that $x^2-y^2$ (B$_{1g}$) shear strain can produce a significant increase in $T_c$, with a small decrease in $T_\text{trsb}$ \cite{grinenko2021split} - i.e. a split transition - while hydrostatic pressure (which produces A$_{1g}$ strain) can produce a pronounced depression of $T_c$ but no detectable splitting of the transition \cite{newhicks}.
These  observations are trivially accounted for if one assumes that $\alpha_3$ has a strong (albeit quadratic) dependence on shear  strain but only weakly dependent on isotropic strain while $\alpha_0$ depends on both components of the strain.

How stringent a condition this places on the isotropic strain dependence of $\alpha_3$ depends on the magnitude and character of the strain inhomogeneities - i.e. the width of the SC transition.
To the extent that we can ignore the effect of $xy$ (B$_{2g}$) shear strain (i.e. for $\alpha_1\approx 0$), it follows that so long as there are regions of $d$ and regions of $g$ wave SC, there must be domain walls between them at which TRS breaking can arise.  Thus,  a spilt transition will be apparent only when the mean value of $\alpha_3$ is greater than its variance.   

 The present considerations are encouraging in the sense that they illustrate  a plausible explanation of a set of previously puzzling experiments in SRO.  However, it is important to reiterate that this analysis sheds no insight of what is probably the most vexing aspect of the proposed scenario: why are these two symmetry distinct forms of SC order  nearly degenerate with one another without need of any fine tuning?

\begin{acknowledgments}
We acknowledge extremely helpful discussions with Clifford Hicks, Catherine Kallin, Tony Leggett, Andy MacKenzie, and Brad Ramshaw.
SAK and AY were supported, in part,  by NSF grant No.  DMR-2000987 at Stanford. EB was supported by the European Research Council (ERC) under grant HQMAT (Grant Agreement No. 817799), the Israel-US Binational Science Foundation (BSF), and a Research grant from Irving and Cherna Moskowitz.
\end{acknowledgments}

\bibliography{main.bib}
\onecolumngrid
\section{Appendix}
\twocolumngrid
\subsection{Derivation of nonlinear sigma model}\label{sec:nonlinear-sigma-derivation}
Here, we review an efficient method of deriving the nonlinear sigma model in Eq. \eqref{eq:nonlinear-sigma}.
Suppose that the OP $\Delta =\Delta(q)$ depends smoothly on parameter $q$. Then it's change $\Delta(q) \to \Delta(q+\delta q)$ can be expressed as an infinitesimal rotation along with a change in the Berry phase $\gamma$, i.e.,
\begin{equation}
    \Delta (q+\delta q) = e^{i \delta \gamma } U(q)^{\delta q} \Delta(q)
\end{equation}
Where $U(q)^{\delta q} \equiv \exp(-iH(q)\delta q)\in SU(2)$ is the infinitesimal rotation with
\begin{equation}
    H(q) \delta q = \frac12 (\hatb{n} \times \delta \hatb{n})\cdot \bm\tau
\end{equation}
Therefore, we can write
\begin{align}
    (-i\partial_\mu -A_\mu)\Delta &= \left( \partial_\mu \chi + a_\mu - A_\mu -H_\mu \right)\Delta \\
    H_\mu &= \frac{1}{2} (\hatb{n} \times \partial_\mu \hatb{n})\cdot \bm{\tau}
\end{align}
Using the (anti-)commutation relations of the Pauli matrices, we can then efficiently derive the nonlinear sigma model, e.g.,
\begin{align}
    \Delta^\dagger H_\mu^2\Delta &= \frac{|\Delta|^2}{2} \tr\left[ H^2_\mu \left(\tau_0 + \hatb{n} \cdot \bm{\tau} \right) \right] \\
    &= |\Delta|^2 \left| \frac{\partial_\mu \hatb{n}}{2}\right|^2
\end{align}
Similarly, we have
\begin{align}
    \Delta^\dagger H_\mu \Delta &= 0 \\
    \Re \left[\Delta^\dagger H_x \tau_1 H_y \Delta\right] &= - n_1 \left(\frac{\partial_x \hatb{n}}{2} \cdot \frac{\partial_y \hatb{n}}{2} \right) \\
    \Re \left[\Delta^\dagger H_\mu \tau_1 \Delta\right] &= \frac{1}{2} [\hatb{n} \times \partial_x \hatb{n}]_1
\end{align}
\subsection{General ground state}\label{sec:general-ground-state}
Here, we present the exact solution of the ground state of the general Ginzburg-Landau free energy in the presence of a uniform strain vector $\bm{\alpha}$.
To provide a more intuitive picture of the general ground state, this section first treats the less restrictive case where $\beta_1\ne \beta_3$ but $\beta_3'=0$.
We then present the ground state in full generality.

It should be noted, however, that the uniform ground state does not depend on the kinetic terms $K$ and thus we can always renormalize $D,G$ so that $\beta_3' =0$.
\subsubsection{$\beta_1\ne \beta_3$ and $\beta_3'=0$}
In the case where $\beta_1 \ne \beta_3$ are positive and $\beta_3'=0$, the potential term of the Ginzburg-Landau free energy can be rewritten as
\begin{align}
  V_2 &= -|\Delta|^2 (\alpha_0 + \alpha_1 n_1+\alpha_3 n_3) \\
  V_4 &= \frac{1}{2} |\Delta|^4 (1 +\beta_1 [n_1]^2 +\beta_3 [n_3]^2 )
\end{align}
Where $\hatb{n}$ is the Bloch vector corresponding to the OP $\Delta$.
It's then clear that the potential $V$ is a strongly joint-convex function of $|\Delta|^2,|\Delta|^2 n_3$ and $|\Delta|^2 n_1$.
Therefore, convex optimization guarantees that the ground state of the uniform system is unique and given by
\begin{equation}
  |\Delta|^2 = \alpha_0,\quad n_i =\frac{\alpha_i}{\beta_i \alpha_0}
\end{equation}
Provided that $[n_1]^2 +[n_3]^2 \le 1$.
Conversely, in the case where the above solution is not feasible, i.e., $[n_1]^2 +[n_3]^2 > 1$, the unique ground state is given by
\begin{align}
  |\Delta|^2 &= \frac{\alpha_0}{1-\lambda} \\
  |\Delta|^2 n_i  &= \frac{\alpha_i}{\beta_i +\lambda}
\end{align}
Where $\lambda\in [0,1)$ is chosen so that $[n_1]^2 +[n_3]^2 =1$.
In particular, the ground state corresponds to a Bloch vector $\hatb{n}$ which is in the $\hatb e_1$-$\hatb e_3$ plane and thus does not break TRS.
\subsubsection{Full generality: V is stable and superconducting}
In full generality, we still require that the Ginzburg-Landau energy is stable, i.e., the potential term $V\to \infty$ in the limit where $|\Delta| \to \infty$.
This corresponds to the condition
\begin{equation}
  \label{eq:stability-condition}
  4\beta_3 -\beta_3'^2 > 0
\end{equation}
Similarly, we are only concerned with nontrivial superconducting ground states, which corresponds to the condition
\begin{equation}
  \label{eq:sc-condition}
  2\alpha_0 \beta_3 -\alpha_3\beta_3' >0
\end{equation}
Therefore, our general ground state solution is subject to the 2 conditions above.
The potential term of the Ginzburg-Landau free energy can be rewritten as
\begin{align}
  V_2 &= -|\Delta|^2 (\alpha_0 +\alpha_1 n_1+\alpha_3 n_3 ) \\
  V_4 &= \frac{1}{2} |\Delta|^4 (1 +\beta_1 [n_1]^2 +\beta_3 [n_3]^2  +\beta_3' n_3)
\end{align}
By our stability condition \eqref{eq:stability-condition}, the potential term $V$ is a strongly joint-convex function in terms of $|\Delta|^2,|\Delta|^2 n_1$ and $|\Delta|^2 n_3$.
Therefore, convex optimization guarantees that the ground state of the uniform system is unique and given by
\begin{align}
  |\Delta|^2 &= 2\times \frac{2\alpha_0 \beta_3 -\alpha_3\beta_3'}{4\beta_3 -\beta_3'^2} >0 \\
  |\Delta|^2 n_3  &= 2\times\frac{2\alpha_3 - \beta_3'\alpha_0}{4\beta_3 -\beta_3'^2} \\
  |\Delta|^2 n_1 &= \frac{\alpha_1}{\beta_1}
\end{align}
Provided that $[n_1]^2 +[n_3]^2 \le 1$.
Conversely, in the case where the above solution is not feasible, i.e., $[n_1]^2 +[n_3]^2 > 1$, the unique ground state is given by
\begin{align}
  |\Delta|^2 &= \frac{2}{4(\beta_3 +\lambda) -\beta_3'^2} \\
  &\times \left[2(\beta_3 +\lambda)\alpha_0 -\alpha_3\beta_3'^2 +4\alpha_3 \frac{(\beta_3+\lambda)\lambda}{\beta_3'}\right]  \nonumber\\
  |\Delta|^2 n_3  &= 2\times \frac{2(1-\lambda)\alpha_3 -\beta_3' \alpha_0}{4(\beta_3 +\lambda) -\beta_3'^2}\\
  |\Delta|^2 n_1 &= \frac{\alpha_1}{\beta_1 +\lambda}
\end{align}
Where $\lambda \ge 0$ is chosen so that $[n_1]^2 +[n_3]^2  =1$.
In particular, the ground state corresponds to a Bloch vector $\hatb{n}$ which is in the $\hatb e_1$-$\hatb e_3$ plane and thus does not break TRS.
\subsection{Narrow Domain Wall Calculations}
\label{sec:narrow-dw-calc}
In this section, we present detailed calculations of a narrow domain wall between the transition of the TRS states $d-g \to d+g$.
To be more concrete, let us consider the case where $\alpha_3(y) = \ubar{\alpha}_3$ is constant, while $\alpha_1(y) = \ubar{\alpha}_1 \sign(y)$ changes sign as $y=-\infty \to +\infty$.
The quartic terms are set to $\beta_1=\beta_3 =\beta >0,\beta_3'=0$ as before and the strain vector magnitude satisfies $\ubar{\alpha} \ge \beta \alpha_0$ so that the uniform ground state at $y\to \pm \infty$ is denoted by a Bloch vector $\hat{\bm{n}}$ pointing in the same direction as the strain vector $\ubar{\bm{\alpha}}$ and that the magnitude $|\Delta|^2 =\alpha_0$.
Without loss of generality, we shall also assume that $\ubar{\alpha}_1,\ubar{\alpha}_3 >0$ so that the angle $\psi$ of the strain vector $\ubar{\bm{\alpha}}\equiv \ubar{\alpha} (\sin \psi,0,\cos \psi)$ satisfies $0< \psi <\pi/2$ as shown in Fig. \eqref{fig:Bloch-domain-walls}.

We then calculate the domain wall energy $\Delta F$ of possible transition paths as illustrated by the blue and green arrows in Fig. \eqref{fig:Bloch-domain-walls} where the magnitude $|\Delta|^2$ of the OP $\Delta$ is assumed to be constant and $=\alpha_0$.
In particular, the blue arrow represents a TRS preserving transition, while the green arrow denotes a TRSB transition restricted to a 2D plane at angle $\omega$ with respect to the $\hatb{e}_3$-axis.

\subsubsection{TRS preserving transition}
We shall first consider a TRS preserving transition as described by the blue path in Fig. \eqref{fig:Bloch-domain-walls}.
Notice that we implicitly assumed that $\psi < \pi/2$.
If on the other hand, $\psi >\pi/2$, then we would consider the complement path wrapping from underneath the Bloch sphere.
Since the transition preserves TRS so that the Bloch vector is in the $\hatb e_1$-$\hatb e_3$ plane, we can restrict the azimuthal angle $\phi \equiv 0$ for $y\ge0$ and $\phi\equiv \pi$ for $y < 0$.
Therefore, we can rewrite the Ginzburg-Landau free energy in terms of polar angle $\theta$ and average phase $\chi$, i.e., if $y\ge 0$ so that $\theta \ge 0$, then
\begin{align}
  \label{eq:GL-TRS}
  F &= V_2 +V_4 +K \\
  V_2 &= -|\Delta|^2 (\alpha_0  +\ubar{\alpha} \cos{(\theta - \psi)}) \\
  V_4 &= +\frac{1}{2}|\Delta|^4 (1+\beta) \\
  K &= \frac{\kappa}{2} |\Delta|^2 \left[ \dot\chi^2 +\left( \frac{\dot \theta}{2}\right)^2\right]
\end{align}
It's then clear that $\chi = \text{const}$ and that
\begin{align}
  \frac{\kappa}{2} \frac{\ddot{\theta}}{2} &= \ubar{\alpha} \sin (\theta -\psi) \\
  \dot{\theta}^2 +\frac{8\ubar{\alpha}}{\kappa} \cos(\theta -\psi) &= \text{const}
\end{align}
Using the boundary conditions $\theta \to \psi$ and $\dot{\theta}\to 0$ as $y\to \infty$ and that $\theta=0$ at $y=0$, we see that
\begin{align}
  \dot{\theta}^2 &= \frac{8\ubar{\alpha}}{\kappa} (1- \cos(\theta -\psi)) \\
  \frac{\dot{\theta}}{2} &= -\sqrt{\frac{4\ubar{\alpha}}{\kappa}} \sin \left( \frac{\theta -\psi}{2}\right) \\
  \theta &= \psi - 4 \arctan \left( \tan{\left(\frac{\psi}{4}\right)} e^{-y/\xi} \right)
\end{align}
Where $\xi = \sqrt{\kappa/ 4\ubar{\alpha}}$ is the characteristic length of the domain wall.
We can similar solve for the transition in the case where $y\le0$ so that in general,
\begin{align}
  \theta &= \left[ \psi - 4 \arctan \left( \tan{\left(\frac{\psi}{4}\right)} e^{-|y|/\xi} \right)\right] \\
  \phi &= \frac{\pi}{2} (1 -\sign y)
\end{align}

We can then calculate the domain wall energy $\Delta F = F-F_0$ for $y\ge 0$ where $F_0 =F[\psi]$ is the Ginzburg-Landau free energy of the uniform ground state, i.e.,
\begin{align}
  \Delta F_\text{trs} &= \int_0^{+\infty} [F[\chi,\theta,\phi] -F_0]dy \\
  &= \kappa \alpha_0 \int_0^{+\infty} \left( \frac{\dot \theta}{2} \right)^2 dy\\
  &= \frac{\kappa \alpha_0}{\xi} \int_0^{\psi} -\sin \left( \frac{\theta-\psi}{2} \right) \frac{d\theta}{2} \\
  &= \frac{2\kappa \alpha_0}{\xi} \sin^2\left(\frac{\psi}{4}\right)
\end{align}
\subsubsection{TRSB transition}
Let us now consider the special TRSB transition as described by the path with $\omega=\pi/2$ in Fig. \eqref{fig:Bloch-domain-walls}, so that the polar angle $\theta$ of the Bloch vector $\hatb{n}$ remains constants and $=\psi <\pi/2$, while the azimuthal angle $\phi$ twists $\pi \to 0$.
Therefore, we can rewrite the Ginzburg-Landau free energy in terms of azimuthal angle $\phi$ and average phase $\chi$, i.e., if $y\ge 0$ so that $0\le \phi \le \pi/2$, then
\begin{align}
  \label{eq:GL-TRSB}
  F &= V_2 +V_4 +K \\
  V_2 &= -|\Delta|^2 (\alpha_0  +\alpha_1 \sin \psi \cos \phi +\alpha_3 \cos \psi ) \\
  V_4 &= +\frac{1}{2} |\Delta|^4 (1 +\beta (1 -\sin^2 \psi \sin^2 \phi)) \\
  K &= \frac{\kappa}{2} |\Delta|^2 \left[ \dot \chi^2 +\left( \frac{\dot \phi}{2}\right)^2 -  \dot{\chi}\dot{\phi}\cos \psi \right]
\end{align}
By the Euler-Lagrange equations, it's then clear that if the average phase $\chi$ satisfies the initial condition $\chi(y\to-\infty) =0$, then
\begin{equation}
  \chi = \frac{\phi-\pi}{2} \cos{\psi}
\end{equation}
In particular, the average phase $\chi$ changes by $\Delta \chi = -(\pi /2)\cos \psi$ as $\phi$ twists $\pi \to0$.
We can similarly solve for $\phi$ using the boundary condition $\dot{\phi},\phi\to 0$ as $y\to +\infty$ so that
\begin{align}
  \frac{\kappa}{2} \frac{\ddot{\phi}}{2} &=  \sin \phi (\ubar\alpha -\beta \alpha_0 \cos{\phi}) \\
  \frac{\dot{\phi}}{2} &= -\frac{1}{\xi} \sin{\left(\frac{\phi}{2}\right)} \sqrt{ 1 -\epsilon \cos^2{\left(\frac{\phi}{2}\right)}}
\end{align}
Where $\epsilon = \beta \alpha_0/\ubar{\alpha}$ and $\xi = \sqrt{\kappa/ 4\ubar{\alpha}}$ is the characteristic length of the domain wall.
Notice that by our ground state discussion \eqref{eq:ground-state-2}, the finite strain $\ubar{\alpha}$ satisfies $\epsilon \le 1$ and thus the square root is well-defined.
We can then solve the first order differential equation so that
\begin{equation}
  \phi = 2 \arccos{ \left( \frac{1-\eta}{\sqrt{(1+\eta)^2 -4\eta \epsilon}}\right)}
\end{equation}
Where
\begin{align}
  \eta &= C \exp{\left(-y \frac{\sqrt{1-\epsilon}}{2\xi} \right)} \\
  \frac{\pi}{2} &= \phi (\eta = C)
\end{align}

We can then calculate the domain wall energy $\Delta F = F-F_0$ for $y\ge 0$ where $F_0 =F[\psi]$ is the Ginzburg-Landau free energy of the uniform ground state, i.e.,
\begin{align}
  \Delta F_\text{trsb} &= \int_0^{+\infty} [F[\chi,\theta, \phi] -F_0]dy \\
  &= \kappa \alpha_0 \sin^2\psi \int_0^{+\infty} \left( \frac{\dot\phi}{2} \right)^2 dy \\
  &= \frac{\kappa \alpha_0}{\xi} \sin^2 \psi \\
  &\times \int_0^{\pi/2} \sin{\left(\frac{\phi}{2}\right)} \sqrt{ 1 -\epsilon \cos^2{\left(\frac{\phi}{2}\right)}} \frac{d\phi}{2}
\end{align}
This can be solved analytically if necessary with the substitution $x = \cos{(\phi/2)}$. In particular, in the extreme large strain limit $\ubar{\alpha} \gg \beta \alpha_0$, the domain wall energy is approximated by
\begin{equation}
  \Delta F  \approx \frac{2\kappa \alpha_0}{\xi} \sin^2{\psi} \sin^2 \left(\frac{\pi}{8} \right)
\end{equation}
\subsubsection{General TRSB transition}
\begin{figure}
\centering
\includegraphics[width=1\columnwidth]{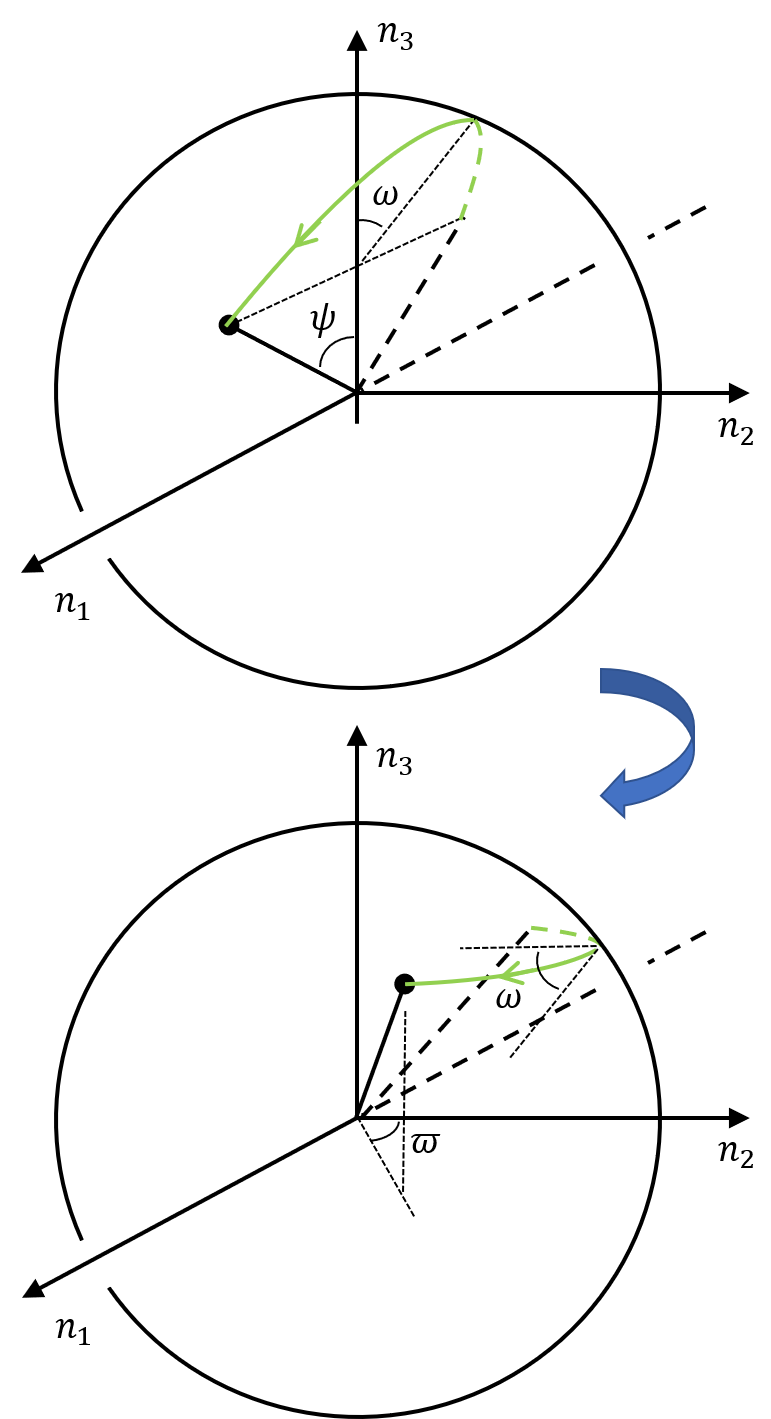}
\caption{General TRSB Domain Wall.
The top figure is the Bloch sphere representation of a general TRSB transition through the domain wall $\alpha_1(y)=\ubar{\alpha}_1 \sign(y)$.
The green arrow which representing the transition path is assumed to be in a plane at angle $\omega$ with respect to the $\hatb{e}_1$-$\hatb{e}_3$ plane.
The bottom figure is the transition rotated of angle $\omega$ about the $+x$-axis in the clockwise direction.
}\label{fig:Bloch-general-dw}
\end{figure}

Let us finally consider a general TRSB transition as described by the green path in Fig. \eqref{fig:Bloch-domain-walls}.
To simplify the problem, we will only consider general TRSB transitions which occur in a 2D plane as shown in Fig. \eqref{fig:Bloch-general-dw}, where the transition plane is at an arbitrary angle $\omega$ with respect to the vertical plane.
If $\omega = 0$, then the transition corresponds to the TRS preserving transition (blue arrow), and if $\omega=\pi/2$, then it corresponds to the special TRSB transition.
Since the Ginzburg-Landau free energy has a global $SU(2)$-symmetry, we can rotate the system about the $+x$-axis of angle $\omega$ in the clockwise direction so that the transition occurs in a plane parallel to $xy$-plane and thus can be parametrized in terms of azimuthal angle $\phi$, while the polar angle $\theta$ is held fixed, as shown in Fig. \eqref{fig:Bloch-general-dw}.
This corresponds to using the rotated order parameter $\tilde{\Delta}=U\Delta$ where $U \in SU(2)$ corresponds to the described rotation $R$.
We then rewrite the Ginzburg-Landau free energy so that if $\utilde{\bm{\alpha}}=R\ubar{\bm{\alpha}}$ is the rotated strain vector and $\vphi \equiv \pi/2 -\phi$ is the azimuthal angle relative to the $+y$-axis, then
\begin{align}
  F &= V_2 +V_4 +K \\
  V_2 &= -|\Delta|^2 (\alpha_0  + \ubar\alpha \cos^2{\theta}) \\
  & -|\Delta|^2 \ubar \alpha \sin^2{\theta} \cos\left(\vphi -\vpi\right) \nonumber \\
  V_4 &= +\frac{1}{2}|\Delta|^4 (1 +\beta ) \\
  &- \frac{1}{2}|\Delta|^4 \beta \left(\frac{\sin{\theta} \cos{\theta}}{\cos{\psi}}\right)^2 \left(\cos \vphi -\cos \vpi \right)^2 \nonumber \\
  K &= \frac{\kappa}{2}|\Delta|^2 \left[ \dot\chi^2 +\left( \frac{\dot \vphi}{2}\right)^2 + \dot \chi \dot \vphi \cos \theta   \right]
\end{align}
Here $\theta$ is held fixed and $\phi$ transitions in a manner such that
\begin{equation}
    \vphi \equiv \frac{\pi}{2}-\phi  = -\vpi \to 0 \to \vpi
\end{equation}
As $y$ goes from $-\infty \to 0\to +\infty$, where $\theta,\vpi$ are determined by the boundary condition $\utilde{\bm{\alpha}} \propto \hat{\bm{n}}(y\to +\infty)$, i.e.,
\begin{align}
  \sin {\psi} &= \sin \theta \sin \vpi \\
  \cos {\psi} \cos \omega &= \sin \theta \cos \vpi \\
  \cos {\psi} \sin{\omega} &= \cos \theta
\end{align}
In particular, $\theta = \pi/2, \vpi = \psi$ for the TRS preserving transition, and $\theta =\psi,\vpi=\pi/2$ for the special TRSB transition.

By the Euler-Lagrange equations, it's then clear that if the global phase $\chi$ satisfies the initial condition $\chi(y\to-\infty) =0$, then
\begin{equation}
  \label{eq:dw-chi}
  \chi = \frac{\phi}{2} \cos{\theta} - \frac{1}{2} \left(\frac{\pi}{2} +\vpi \right) \cos{\theta}
\end{equation}
Hence, the average phase $\chi$ changes by $\Delta \chi = -\vpi \cos \theta $ as $y=-\infty \to \infty$.
We can similarly solve for $\vphi$ so that
\begin{align}
  \frac{\dot{\vphi}}{2} &= -\frac{1}{\xi} \sin \left( \frac{\vphi -\vpi}{2}\right) \\
  &\times \sqrt{1 -(\epsilon \sin^2 \omega )\sin^2 \left( \frac{\vphi -\vpi}{2} +\vpi\right) } \nonumber
\end{align}
Where $\epsilon = \beta \alpha_0/\ubar{\alpha} \le 1$ and $\xi = \sqrt{\kappa/ 4\ubar{\alpha}}$ is the characteristic length of the domain wall.
It should be noted that the equation can be solved analytically for arbitrary $\epsilon \le 1$, albeit in implicit form, i.e., $f(\vphi) = y$ for some function $f$.
It should be noted that the explicit form $\vphi = \vphi(y)$ is not necessary to compute the domain wall energy $\Delta F$.
Indeed, we have
\begin{align}
  \Delta F &= \int_0^{+\infty} [F[\chi,\theta,\phi] -F_0]dy \\
  &= \kappa \alpha_0 \sin^2 \theta \int_0^{+\infty} \left( \frac{\dot \phi}{2} \right)^2  dy \\
  &= \frac{\kappa \alpha_0}{\xi} \sin^2 \theta  \\
  & \times \int_0^{\vpi/2} \sin x \sqrt{1-(\epsilon \sin^2 \omega ) \sin^2 \left( \vpi -x\right) } dx \nonumber
\end{align}
In the extreme large strain limit $\ubar{\alpha} \gg \beta \alpha_0$, the domain wall energy is approximated by
\begin{equation}
  \Delta F  \approx \frac{2\kappa \alpha_0}{\xi} \sin^2{\theta} \sin^2 \left(\frac{\vpi}{4} \right)
\end{equation}

\subsection{Full self-consistent equations}
\label{sec:BCS-calc}
In this section, we derive the full self-consistency field equations (SCFs) for the full Hamiltonian $H_\text{full}=H_0 +H_1$ defined in Eq. \eqref{eq:H0}, \eqref{eq:H1}. Let us first write the corresponding BCS Hamiltonian $H$ as
\begin{align}
H &= \sum_{\rvec', \rvec, s} \cT(\rvec',\rvec) c_{\rvec's}^\dagger  c_{\rvec s}^{\vphantom{\dagger}} \\
&+ \sum_{\rvec',\rvec} (\Delta_{\rvec',\rvec} c_{\rvec'\uparrow}^\dagger  c_{\rvec\downarrow}^\dagger + \text{h.c.}) \\
&=
 \begin{bmatrix}
   c_{\uparrow}^\dagger & c_{\downarrow}
 \end{bmatrix}
 \hat{H}
 \begin{bmatrix}
   c_{\uparrow} \\
   c_{\downarrow}^\dagger
 \end{bmatrix}, \qquad
 \hat{H} =
 \begin{bmatrix}
   \cT & \Delta \\
   \Delta^\dagger & -\bar{\cT}
 \end{bmatrix}
\end{align}
To simplify notation, let us introduce the notation $1_{\rvec} = |\rvec\ket\bra \rvec|$ for the projection operator and the 1-particle density matrices (\textbf{1-pdms}) \cite{bach1994generalized} of the BCS Hamiltonian $H$, i.e.,
\begin{align}
\label{eq:pdm}
\gamma(\rvec',\rvec) &= \bra c_{\rvec\uparrow}^\dagger c_{\rvec'\uparrow} \ket = \bra c_{\rvec\downarrow}^\dagger c_{\rvec'\downarrow} \ket \\
\alpha(\rvec',\rvec) &= \bra c_{\rvec\downarrow} c_{\rvec'\uparrow} \ket = -\bra c_{\rvec\uparrow} c_{\rvec'\downarrow} \ket
\end{align}
The Hartree-Fock energy $\bra H_\text{full}\ket =\bra H_0\ket +\bra H_1\ket$ can then be computed via Wick's theorem so that
\begin{align}
\label{eq:HF-energy}
\bra H_0 \ket &= 2 \tr (\cT \gamma) \\
\bra H_1 \ket &= -2 \sum_{\tau =d,g} \lambda_\tau \sum_{\rvec} \tr \left(2 f_\tau \alpha^\dagger 1_{\rvec} \alpha f_\tau^{\dagger} 1_{\rvec} \right.\\
&+ \left.f_\tau^{\dagger} \gamma 1_{\rvec} \gamma f_\tau 1_{\rvec} +\gamma 1_{\rvec} f_\tau^{\dagger} \gamma f_\tau 1_{\rvec}\right) \nonumber
\end{align}
The self-consistency field equations (SCFs) can then be derived from extremizing the variational free energy $\delta F \equiv \delta \bra H_\text{full}\ket -T\delta S_H=0$ and written in operator form
\begin{align}
\label{eq:SCF}
\cT &= t -\sum_{\tau=d,g} \lambda_\tau \sum_{\rvec \in \Lambda} \left(( f_\tau  1_{\rvec} f_\tau^\dagger \gamma 1_{\rvec} + \text{h.c.}) \right. \\
&\qquad\qquad\quad \left.+1_{\rvec} f_\tau^\dagger \gamma f_\tau 1_{\rvec} + f_\tau 1_{\rvec} \gamma 1_{\rvec} f_\tau^\dagger\right) \nonumber \\
\Delta &= -2 \sum_{\tau=d,g} \lambda_\tau \sum_{\rvec} (f_\tau 1_{\rvec} \alpha f_\tau^\dagger 1_{\rvec} + \text{tr})
\end{align}
Where $\text{tr}$ implies the transpose term $=1_{\rvec} f_\tau^\dagger \alpha 1_{\rvec} f_\tau = 1_{\rvec} \alpha f_\tau^\dagger 1_{\rvec} f_\tau$, and we used the fact that $\alpha,f_\tau$ are symmetric, i.e., $\alpha^T=\alpha,f_\tau^T=f_\tau$.
The proposed format of writing the SCFs in operator form has the advantage of working in any basis set, and in particular, we can work in real space $|\rvec\ket$ or momentum space $|\vec{k}\ket$.
This will simplify our algebra later on.
The corresponding BdG equations can similarly be written as
\begin{align}
\label{eq:BdG}
\gamma &= u n_E u^\dagger +\bar{v} (1-n_E) v^T \\
\alpha &= -\frac{1}{2} \left(u \tanh\left(\frac{\beta E}{2}\right) v^\dagger +\text{transpose} \right)
\end{align}
where $n_E$ is the diagonal matrix with entries of the Fermi distribution $n_E = (e^{\beta E} +1)^{-1}$ and $u,v$ are chosen so that the following unitary matrix $W$ diagonalizes the first quantized Hamiltonian $\hat{H}$, i.e.,
\begin{equation}
\label{eq:bogoliubov}
\hat{H} = W
\begin{bmatrix}
  E & 0 \\
  0 & -E
\end{bmatrix}
W^\dagger,\quad W =
\begin{bmatrix}
  u & -\bar{v} \\
  v & \bar{u}
\end{bmatrix}
\end{equation}
Notice that the BdG equations \eqref{eq:BdG} can also be written in the more compact form \cite{bach1994generalized}
\begin{equation}
\Gamma = W
\begin{bmatrix}
  n_E & 0 \\
  0 & 1-n_E
\end{bmatrix} W^\dagger
\end{equation}
Where
\begin{equation}
\Gamma =
\begin{bmatrix}
  \gamma & \alpha \\
  \alpha^\dagger & 1-\bar{\gamma}
\end{bmatrix}
= \frac{1}{e^{\beta \hat{H}} +1}
\end{equation}
\subsubsection{Uniform system in momentum space}
The full SCFs can be further simplified in a uniform system, so that the 1-pdms $\gamma,\alpha$ \eqref{eq:pdm}, hopping matrix $t$ and form factors $f_\tau$ are diagonalized in $\vec{k}$-space.
In particular,
\begin{align}
  \label{eq:pdm-k}
  \gamma(\vec{k}) &= \bra c_{\vec{k}\uparrow}^\dagger c_{\vec{k}\uparrow}^{\vphantom{\dagger}} \ket = \bra c_{\vec{k}\downarrow}^\dagger c_{\vec{k}\downarrow}^{\vphantom{\dagger}} \ket \\
  \alpha(\vec{k}) &= \bra c_{-\vec{k}\downarrow} c_{\vec{k}\uparrow} \ket = -\bra c_{\vec{k}\uparrow} c_{-\vec{k}\downarrow} \ket
\end{align}
In this case, the SCFs \eqref{eq:SCF} are reduced to
\begin{align}
  \label{eq:SCF-k}
  \cT(\vec{k}) &= t(\vec{k}) -\sum_{\tau =d,g} \lambda_\tau \left(f_{\tau}(\vec{k}) \llb f_\tau \gamma\rrb + \text{h.c}) \phantom{\frac12}\right. \\
  &\qquad\qquad\qquad \left.\phantom{\frac12}+ \llb  \gamma |f_\tau|^2 \rrb +|f_\tau(\vec{k})|^2 \llb \gamma \rrb \right)\nonumber\\
  \Delta(\vec{k}) &= -4 \sum_{\tau =d,g} \lambda_\tau f_\tau(\vec{k}) \llb \alpha f_\tau^\dagger \rrb
\end{align}
Where we use the notation $\llb \cdots \rrb$ as the average value summed over $\vec{k}$-space, i.e.,
\begin{equation}
  \llb h \rrb = \int_{(-\pi,\pi]^2} \frac{d^2 \vec{k}}{(2\pi)^2} h(\vec{k})
\end{equation}
Notice that the specific form of our $\Delta$-function \eqref{eq:SCF-k}, i.e., $\Delta(\vec{k}) = D f_d(\vec{k}) + G f_g (\vec{k})$ for complex constants $D,G$ implies that the SCFs indeed yield a two-component OP theory.
Similarly, the BdG equations are reduced to
\begin{align}
  \label{eq:BdG-k}
  \gamma(\vec{k}) &= \frac{1}{2} \left( 1+\frac{\cT(\vec{k})}{E(\vec{k})} \right) n(\vec{k}) \\
  &+\frac{1}{2} \left( 1-\frac{\cT(\vec{k})}{E(\vec{k})} \right) \left(1-n(\vec{k})\right) \nonumber \\
  \alpha(\vec{k}) &= -\frac{\Delta(\vec{k})}{2 E(\vec{k})} \tanh \left( \frac{\beta E(\vec{k})}{2} \right)
\end{align}
Where $n(\vec{k}) = (e^{\beta E(\vec{k})} +1)^{-1}$ is the Fermi distribution of the Bogoliubov quasi-particles with dispersion relation
\begin{equation}
    E(\vec{k}) = \sqrt{\cT(\vec{k})^2 +|\Delta(\vec{k})|^2}
\end{equation}
\subsection{$\nabla\cdot J=0$ in BCS theory}
\label{sec:zero-div-proof}
Let $H[\psi]$ denote the BCS Hamiltonian with parameters $\psi =(\xi,\Delta)$. Let the particular choice of parameter $\vphi$ be such that $H [\vphi]$ satisifes the \textit{full} self-consistency equations with respect to the full Hamiltonian $H_\text{full}$, i.e.,
\begin{equation}
  \left. \frac{\partial}{\partial \psi} \right|_{\psi =\vphi} \bra H_\text{full}\ket_{\psi} = \frac{1}{\beta} \left. \frac{\partial}{\partial \psi} \right|_{\psi =\vphi} S[\psi])
\end{equation}
Where $S[\psi]$ is the von-Neumann entropy of $H[\psi]$ defined by $S = -\tr{(\rho \log{\rho})}$ where $\rho$ is the Gibbs distribution of the BCS Hamiltonian $H[\psi]$, and $\bra\cdots \ket_\psi$ is the thermal average at temperature $T$ with respect to the BCS Hamiltonian $H[\psi]$.

We shall subsequently show that at any temperature and every lattice site $\rvec$, the charge density $\rho(\rvec)$ is constant in BCS theory, i.e.,
\begin{equation}
  \left.\frac{\partial \rho(\rvec)}{\partial t} \right|_{t=0} = 0
\end{equation}
As a corollary, we can apply the continuity equation $\partial_t \rho (\rvec) = \nabla\cdot J(\rvec) \equiv \sum_{\rvec'} J(\rvec', \rvec)$ to obtain
\begin{equation}
  \bra \nabla\cdot J(\rvec)\ket_\vphi \equiv \sum_{\rvec' \in \Lambda} \bra J(\rvec',\rvec) \ket_\vphi = 0
\end{equation}

Indeed, let us introduce the notation for gauge transformation at lattice site $\rvec$, i.e., if $M$ is an operator, e.g., the full Hamiltonian $H_\text{full}$ or the BCS Hamiltonian $H[\psi]$, then define
\begin{equation}
  M(s) = e^{is\rho(\rvec)} A e^{-is \rho(\rvec)}, \quad s\in \bR
\end{equation}
In this case, notice that
\begin{equation}
  H[\vphi](s) = H [\vphi(s)]
\end{equation}
where $\vphi(s)$ denote the parameters
\begin{align}
  \cT(\rvec'',\rvec')(s) &= \cT(\rvec'',\rvec') e^{is (\delta(\rvec'',\rvec) -\delta(\rvec',\rvec))}\\
  \Delta(\rvec'',\rvec')(s) &= \Delta(\rvec'',\rvec') e^{is (\delta(\rvec'',\rvec) +\delta(\rvec',\rvec))}
\end{align}

Also notice that the BCS partition function $Z [\vphi (s)]$ is independent of $s$ since the trace is invariant under unitary gauge transforms and thus
\begin{align}
  &\left\bra \frac{\partial H_\text{full}(s)}{\partial s} \right\ket_{\vphi_s}  \nonumber \\
  &\qquad= \frac{1}{Z_\vphi} \tr \left( e^{-\beta H [\vphi_s]} \frac{\partial H_\text{full}(s)}{\partial s}\right) \\
  &\qquad= +\frac{1}{Z_\vphi} \frac{\partial}{\partial s} \tr \left( e^{-\beta H [\vphi_s]} H_\text{full}(s)\right)  \\
  &\qquad\quad- \frac{1}{Z_\vphi} \tr \left( H_\text{full}(s) \frac{\partial}{\partial s} e^{-\beta H [\vphi_s]} \right)
\end{align}
The first term is $=0$ since the trace is invariant under unitary transforms. Setting $s=0$, we see that
\begin{align}
  &-\left\bra \left.\frac{\partial H_\text{full}(s)}{\partial s}\right|_{s=0} \right\ket_\vphi \nonumber \\
  &\qquad= \frac{1}{Z_\vphi} \tr \left( H_\text{full} \left.\frac{\partial}{\partial s}\right|_{s=0} e^{-\beta H [\vphi_s]} \right) \\
  &\qquad= \left.\frac{\partial}{\partial s}\right|_{s=0} \frac{1}{Z_\vphi} \tr \left( H_\text{full}  e^{-\beta H [\vphi_s]} \right) \\
  &\qquad= \left. \frac{\partial}{\partial s} \right|_{s=0} \bra H_\text{full} \ket_{\vphi_s} \\
  &\qquad= \left. \frac{\partial \vphi_s}{\partial s} \right|_{s=0} \left. \frac{\partial }{\partial \psi} \right|_{\psi=\vphi} \bra H_\text{full}\ket_{\psi} \\
  &\qquad= \frac{1}{\beta} \left. \frac{\partial \vphi_s}{\partial s} \right|_{s=0} \left. \frac{\partial}{\partial \psi} \right|_{\psi =\vphi} S[\psi] \\
  &\qquad= \frac{1}{\beta} \left. \frac{\partial}{\partial s} \right|_{s =0} S[\vphi_s]
\end{align}
Notice that $S[\vphi_s]$ is independent of $s$ since the trace is invariant under unitary gauge transforms. Hence, the right-hand-side is $=0$ and thus we arrive at the statement
\begin{equation}
  \left\bra \left.\frac{\partial \rho(\rvec)}{\partial t} \right|_{t=0} \right\ket_\vphi = -\left\bra \left.\frac{\partial H_\text{full}(s)}{\partial s}\right|_{s=0} \right\ket_\vphi = 0
\end{equation}

\end{document}